\documentclass[12pt,fleqn,letterpaper]{article}
\usepackage{graphicx} 
\usepackage{fullpage}
\usepackage{amssymb}
\usepackage{amsthm}
\usepackage{epstopdf}
\usepackage[flushleft]{threeparttable}
\usepackage{pdflscape}
\usepackage{amsmath}
\usepackage{tkz-graph}
\usetikzlibrary{shapes.geometric}
\usepackage{subfigure}
\usepackage{setspace}
\usepackage[bottom]{footmisc}
\usepackage{endnotes}
\usepackage{graphicx}
\usepackage{textcomp}
\usepackage{adjustbox} 
\usepackage{natbib}
\usepackage[ansinew]{inputenc}
\usepackage{verbatim}
\usepackage{lscape}
\usepackage{array}
\usepackage{fltpoint}
\usepackage{dcolumn}
\usepackage{booktabs}
\usepackage{rotating}
\usepackage[norounding,point]{rccol}
\usepackage{hyperref}
\usepackage{placeins}
\usepackage{multirow}
\usepackage{paralist}
\usepackage{enumitem}
\usepackage{subcaption}

\usepackage{wasysym}
\usepackage{mathtools}
\usepackage{marvosym}
\usepackage{amsthm}

\usepackage[margin=1in]{geometry}

\usepackage{tikz}

\usetikzlibrary{shapes,decorations,arrows,calc,arrows.meta,fit,positioning}
\tikzset{
    -Latex,auto,node distance =1 cm and 1 cm,semithick,
    state/.style ={ellipse, draw, minimum width = 0.7 cm},
    point/.style = {circle, draw, inner sep=0.04cm,fill,node contents={}},
    bidirected/.style={Latex-Latex,dashed},
    el/.style = {inner sep=2pt, align=left, sloped}
}

\numberwithin{equation}{section}

\hypersetup{
colorlinks=true,
urlcolor=blue,
linkcolor=red,
citecolor=blue
}

\usepackage{setspace}
\usepackage{xspace}

\newtheorem{theorem}{Theorem}

\newtheorem{corollary}{Corollary}

\newtheorem{lemma}{Lemma}

\newtheorem{assumption}{Assumption}

\newtheorem{assumptionmod}{Assumption}

\tikzstyle{VertexStyle} = [shape            = ellipse,
minimum width    = 6ex,%
draw]

\tikzstyle{EdgeStyle}   = [->,>=stealth']

\begin{document}
\sloppy

\begin{center}

{\LARGE Testing Identification in Mediation and Dynamic Treatment Models}

{\large \vspace{0.1cm}}

{\large Martin Huber*, Kevin Kloiber+, Luk\'{a}\v{s} Laff\'{e}rs**} \vspace{0.1cm}

{\footnotesize{*Department of Economics, University of Fribourg, Fribourg, Switzerland \\ +Department of Economics, University of Munich, Munich, Germany\\  **Department of Mathematics, Matej Bel University, Bansk\'a Bystrica, Slovakia\\ **Department of Economics, Norwegian School of Economics, Bergen, Norway}}
\end{center}

\thispagestyle{empty}
\begin{abstract}
  {\footnotesize We propose a test for the identification of causal effects in mediation and dynamic treatment models, which is based on two sets of observed variables: covariates to be controlled for and suspected instruments. Building on \citet{huberkueck2022} for single treatment models, we extend the framework to models with a sequential assignment of a treatment and a mediator, allowing for the identification of direct, indirect, or joint treatment effects. We derive testable conditions in observational data, which jointly imply: (1) exogeneity of the treatment and the mediator conditional on covariates, and (2) validity of distinct instruments for the treatment and the mediator, meaning that the instruments do not directly affect the outcome and are unconfounded given the covariates. Our framework also allows joint testing of the selectivity of treatment and attrition by replacing the mediator with a selection indicator for observing the outcome. We propose a machine learning-based test to control for covariates and analyze its finite sample performance in a simulation study. Additionally, we apply our method to Slovak labor market data, finding that our testable implications are not rejected for a sequence of training programs typical in dynamic treatment evaluations.}
  \\[0.5cm]\it{JEL Classification:} C12, C21, C26 \\[0.1cm]
\end{abstract}

\begin{spacing}{1}
{\scriptsize  We have benefited from the comments of seminar participants in Neuch\^atel, D\"usseldorf, and Kassel, as well as conference participants at the European Causal Inference Meeting 2024 in Copenhagen, COMPIE 2024 in Amsterdam and the annual conference of the IAAE 2024 in Thessaloniki. Addresses for correspondence: Martin Huber, University of Fribourg, Bd.\ de P\'{e}rolles 90, 1700 Fribourg, Switzerland; martin.huber@unifr.ch. Kevin Kloiber, University of Munich, Ludwigstrasse 33, 80539 Munich, Germany, and ifo Institute, Poschingerstrasse 5, 81679 Munich, Germany; kevin.kloiber@econ.lmu.de. Luk\'{a}\v{s} Laff\'{e}rs, Department of Mathematics, Matej Bel University, Tajovsk\'{e}ho 40, 974 01 Bansk\'{a} Bystrica, Slovakia; Department of Economics, Norwegian School of Economics, Helleveien 30, 5045 Bergen, Norway; lukas.laffers@gmail.com. Laff\'{e}rs acknowledges support  provided by the Slovak Research and Development Agency under contracts VEGA 1/0398/23 and APVV-21-0360.}
\end{spacing}
%
%
%

\thispagestyle{empty}
\setcounter{page}{1}

\section{Introduction} 

Causal mediation concerns the evaluation of the direct effect of a treatment on an outcome, as well as the indirect effect mediated by an intermediate outcome referred to as a mediator. For instance, when evaluating the impact of two sequential training programs (such as job application training followed by an IT course) on employment, the direct effect refers to the impact of the first program, net of participation in the second program. Conversely, the evaluation of dynamic treatment effects pertains to assessing the effects of specific sequences of treatment and mediator values, such as the impact of participating in two sequential training programs versus not participating in any program. Effect identification in numerous studies within the fields of causal mediation and dynamic treatment evaluation relies on a sequential selection-on-observables or ignorability assumption (see, for instance, \citet{Ro86}, \citet{RoHeBr00}, \citet{Lech09},  \citet{FlFl09}, \citet{ImKeYa10}, \citet{Hong10}, \citet{TchetgenTchetgenShpitser2011}, and \citet{Huber2012}). This assumption postulates that, conditional on observed covariates, the treatment, and the mediator are exogenous, such that their causal effects are not confounded by unobservables.

Similar identification issues also arise when assessing causal effects in the context of sample selection or post-treatment outcome attrition, where the outcome is observed only for a selectively non-random subpopulation. For instance, wages are only observed for employed individuals. In such cases, sequential ignorability means that both the treatment and a selection indicator for the observability of the outcome (rather than a mediator) are exogenous conditional on observed characteristics. This allows for the identification of the total effect of the treatment on the outcome, as discussed, for instance, in \citet{Negi2020} and \citet{bia2023double}. Whether the observed covariates plausibly satisfy the assumption of sequential ignorability is conventionally motivated based on theory, intuition, domain knowledge, or previous empirical findings. However, this assumption is often controversial, as unobserved confounders can rarely be entirely ruled out in empirical applications based on observational data.

This study introduces a test for conditions that imply sequential ignorability, facilitating the verification of identification in observational data. Our testing approach relies on two types of observed variables for each treatment and mediator (or selection indicator): covariates to be controlled for and additional variables presumed to be distinct instruments for the treatment and mediator. The testable condition arises within a specific causal structure that rules out certain forms of reverse causality, such as those from the outcome to the mediator, treatment, covariates, or the suspected instruments, and imposes that the suspected instruments have a (first stage)  association with the treatment or the mediator on other observed variables. Under these circumstances, if the instrument for the treatment is conditionally independent of the outcome, given the treatment and covariates, and the instrument for the mediator is conditionally independent of the outcome, given the treatment, mediator, and covariates, then the following conditions hold:
(A) The instruments are valid, meaning they do not directly influence the outcome, except through their impact on the treatment or mediator, and are not associated with unobservable factors affecting the outcome, given the observed variables. (B) The treatment and the mediator are exogenous, implying that they are not associated with unobservable factors that affect the outcome, conditional on the observed variables.

Thus, the testable conditional independence assumption implies that the treatment and mediator satisfy a sequential ignorability assumption, permitting the identification of dynamic treatment effects or the controlled direct effect, as discussed in \citet{Pearl01}. The controlled direct effect corresponds to the net effect of the treatment when holding the mediator fixed at a specific value (e.g., one). In sample selection models, this assumption allows for the evaluation of the total treatment effect. Furthermore, if it additionally holds that the supposed instrument for the treatment is conditionally independent of the mediator given the treatment and the covariates, the effect of the treatment on the mediator is identifiable. This is also a precondition for the evaluation of natural direct and indirect effects as discussed in \citet{RoGr92} and \citet{Pearl01}, albeit their identification rests on somewhat stronger conditions than those implied by our testable implications. Natural effects are defined based on setting the mediator to its potential value under a specific treatment (rather than fixing it at a constant value like one, as is the case when evaluating the controlled direct effect).

To the best of our knowledge, our study is the first to propose an identification test in the contexts of causal mediation, dynamic treatment effects, or sample selection, building on the work of \citet{huberkueck2022} for testing the identification of the (total) effect of a single treatment variable. We extend the machine learning-based testing approach presented therein to enable testing conditional independence assumptions involving multiple variables, such as the treatment, mediator, and selection indicator. Additionally, our extension allows for the consideration of multivalued instruments for the treatment, mediator, or selection indicator, as opposed to binary instruments. The utilization of machine learning methods offers the advantage of data-driven control for important observed confounders during testing, which is particularly valuable in high-dimensional contexts where numerous potential control variables are available. In essence, we propose a test based on the expectation of squared differences between regression functions that include and exclude the respective instruments. The squared difference satisfies the \citet{Neyman1959}-orthogonality condition, implying that we may account for covariates in the regression functions by machine learning without compromising on desirable asymptotic properties, given that specific regularity conditions hold, see \citet{doubleML}. 

We investigate the finite sample performance of our test in a simulation study and find that it performs very decently in terms of empirical size and power in our simulation designs when the sample size consists of several thousand observations (or more). Furthermore, we apply our method to large administrative data from Slovakia to test the identification of the effects of sequential labor market programs for jobseekers on employment. We use the availability in public employment service centers as instruments and also control for a rich set of jobseeker characteristics. Our findings indicate that our testable implications are not rejected for the specific sequences of programs tested, namely a graduate practice program followed by employment incentives (consisting of hiring incentives and subsidized employment).

Our paper contributes to a growing literature on testing identifying assumptions. For instance, \citet{deLunaJohansson2012},  \citet{BlackJooLaLondeSmithTaylor2015}, and \citet{chen2018testing} consider the same or related testable implications as \citet{huberkueck2022} to assess the conditional exogeneity of a treatment based on a valid instrument, rather than jointly testing the treatment exogeneity and instrument validity. A joint test is also proposed by \citet{angrist2017leveraging} in the context of linear regression models to evaluate the value-added of school choice on students' academic performance, whereas \citet{huberkueck2022} and the current paper focus on nonparametric testing methods with possibly high dimensional covariates. \citet{peters2015causal} employ instruments to learn in a data-driven way which variables are treatments, in the sense that they directly influence an outcome of interest, under the assumption that they satisfy conditional exogeneity. The approach utilizes instruments in a way that enables the rejection of treatments violating conditional exogeneity, thereby providing the power to detect identification failures. \citet{angrist2015wanna} test an analogous implication as in \citet{deLunaJohansson2012}, \citet{BlackJooLaLondeSmithTaylor2015}, and \citet{huberkueck2022}, but when the treatment is unconfounded, to test the validity of an instrument within the framework of the regression discontinuity design (RDD). In a sharp RDD, the treatment is a deterministic function of a cutoff in a running variable and, therefore, satisfies conditional exogeneity by design. This allows for testing whether the running variable is a valid instrument, i.e., whether it is not associated with the outcome conditional on the treatment. If this holds, causal effects can also be identified away from the cutoff, which is otherwise not feasible due to the lack of common support in the treatment across different values of the running variable.

Our paper is also related to the literature on causal discovery, which aims to recover a causal structure consistent with both the data and a model specification, typically through a sequence of conditional independence tests (see, e.g., \citet{scholkopf2021toward}; \cite{guo2020survey}; \citet{huber2024introduction}, for reviews). In contrast, the objective of this paper is not to reconstruct a causal graph (or set of graphs), but rather to establish testable conditions on identification assumptions that are sufficient to identify the causal effects of interest.

The remainder of this paper is organized as follows. Section~\ref{Assumptions1} presents the identifying assumptions and testable implications in scenarios involving only pre-treatment covariates as control variables. Section~\ref{Assumptions2} considers a modified causal framework that allows the first instrument (for the treatment) to directly influence the second instrument (for the mediator), which requires controlling for the second instrument in a specific way when testing. Section~\ref{Assumptions3} discusses a setup with dynamic confounding, which requires controlling for post-treatment covariates in addition to pre-treatment covariates when testing. Section~\ref{testing} outlines the machine learning-based test, which permits controlling for high-dimensional covariates in a data-driven manner.  Section~\ref{simulations} presents a simulation study that investigates the finite sample performance of our test. Section~\ref{empirical} provides an empirical application to Slovak labor market data. Section~\ref{conclusion} concludes. The proofs of the identification results are provided in the Appendix.


\section{Assumptions and implications with pre-treatment covariates}\label{Assumptions1}

In causal mediation analysis, the objective is to decompose the overall causal impact of a treatment variable \( D \) on an outcome variable \( Y \) into two distinct components: the direct effect of the treatment on the outcome and the indirect effect that operates through the mediator variable \( M \). A conceptually related framework is dynamic treatment evaluation, where \( D \) represents an initial treatment and \( M \) a subsequent treatment. This framework evaluates the effectiveness of different treatment sequences involving \( D \) and \( M \). In both mediation and dynamic treatment models, the variables \( D \), \( M \), and \( Y \) can exhibit either discrete or continuous distributions. In models involving sample selection or post-treatment outcome attrition, \( M \) functions as a binary selection indicator that determines the observability of outcome \( Y \), while \( D \) represents the treatment. Despite this distinction, the identification challenges encountered in sample selection models are related to those in dynamic treatment models.

To formalize the assumptions required for identifying causal effects in mediation, dynamic treatment, and sample selection models, we use the potential outcome framework (e.g., \citet{Neyman23} and \citet{Rubin74}) and denote random variables by capital letters and their specific values by lowercase letters. Specifically, we denote by $M(d)$ the potential mediator under treatment value $d \in \mathcal{D}$, where $\mathcal{D}$ is the support of the treatment. $Y(d,m)$ denotes the potential outcomes as a function of both the treatment and some mediator value $m \in \mathcal{M}$, where $\mathcal{M}$ is the support of the mediator. We define $Y(d,m)$ and $M(d)$ as functions of an individual's treatment status $D=d$ and mediator value $M=m$ assuming (i) the individual's potential outcomes are unaffected by the treatment or mediator status of others, and (ii) there are no multiple versions of any treatment or mediator level across individuals. This is known as Stable Unit Treatment Value Assumption (SUTVA), see e.g.\  \citet{Rubin80} and \citet{Cox58}. We denote observed pre-treatment covariates as $X$, instrumental variable(s) for the treatment $D$ as $Z_1$, and instrumental variable(s) for the mediator $M$ as $Z_2$. The properties of these instrumental variables are yet to be established. Finally, let $\mathcal{X}, \mathcal{Z}_1, \mathcal{Z}_2$, and $\mathcal{Y}$ denote the support of $X, Z_1, Z_2$, and $Y$. 

We subsequently discuss a range of identifying assumptions for causal mediation analysis, which permit identifying direct and indirect effects conditional on the covariates and testing identification based on specific assumptions on the instruments. Our first assumption establishes a specific causal structure between variables within our framework, positing that only certain variables exert a causal influence on others. We formalize this causal structure using the previously mentioned potential outcome notation, by applying the latter also to other variables than the outcome and the mediator. 
Furthermore, we assert that if a causal relationship exists between two variables (potentially conditional on other variables), then a statistical dependence must exist between them, consistent with the principle of causal faithfulness. 

\begin{assumption}[Causal structure and faithfulness]\label{A1}
\begin{align*}
  M(y)=M, D(m, z_2, y)=D, X(d, m, z_2, y)=X,  Z_1(d, m, z_2, y)=Z_1,\; \textrm{ and }   Z_2(m, y)=Z_2,  \\ 
  \forall d \in \mathcal{D}, m \in \mathcal{M},   z_2 \in \mathcal{Z}_2  \textrm{ and }  y  \in \mathcal{Y},\\
\end{align*}
  \\[-2cm]only variables which are d-separated in any causal model are statistically independent.
\end{assumption}

The first line of Assumption~\ref{A1} excludes any reverse causal effect of outcome $Y$ on $D$, $X$, $M$, $Z_1$, or $Z_2$. In addition, the treatment $D$ must not causally affect $X$, $Z_1$, the mediator $M$ must not causally affect $X$, $D$, $Z_1$, $Z_2$, while $X$ might affect $D$, $M$, $Z_1$, $Z_2$ or $Y$, $Z_1$ might affect $D$. $Z_2$ might affect $M$. The second line of Assumption~\ref{A1} establishes causal faithfulness, which ensures that only variables that are d-separated, i.e., not linked via any causal paths, are statistically independent or conditionally independent, see e.g. \citet{Pearl00}. To be more precise, we employ the d-separation criterion of \citet{pearl1988probabilistic}, which relies on blocking causal paths between variables. A path between two sets of variables, $A$ and $B$, is blocked when conditioning on a set of control variables, $C$, if: 
\begin{enumerate}
	\item the path between $A$ and $B$ is a causal chain, implying that $A\rightarrow M \rightarrow B$ or $A\leftarrow M \leftarrow B$, or a confounding association, implying that $A\leftarrow M \rightarrow B$, and variable (set) $M$ is among control variables $C$ (i.e.\ controlled for),
	\item the path between $A$ and $B$ contains a collider, implying that $A\rightarrow  S  \leftarrow B$, and variable (set) $S$ or any variable (set) causally affected by $S$ is not among control variables $C$ (i.e.\ not controlled for).
\end{enumerate}
According to the d-separation criterion, $A$ and $B$ are d-separated when conditioning on a set of control variables $C$ if, and only if, $C$ blocks every path between $A$ and $B$.\ d-separation is sufficient for the (conditional) independence of two variables, providing a vital component for the proof of our Theorems. Causal faithfulness imposes that d-separation is also a necessary condition, such that two variables are statistically independent if and only if d-separation holds.
A scenario in which faithfulness fails is that one variable affects another one via several causal paths (or mechanisms) which exactly cancel out such that the variables are independent, see e.g.\ the discussion in \citet{spirtes2000causation}.

Assumption~\ref{A1} restricts the set of causal structures that we consider. We refer to the SWIG framework of \citet{richardson2013single} that provides a principled way to read off conditional independence statements about potential outcomes directly from the causal graph and also to label the variables coherently (such as when change $Y$ to $Y(d)$, $Y(m)$ or $Y(d,m)$ in the interventional graph). 

Next, we introduce two common support assumptions that are required for nonparametric identification and testing. To this end, let $f(A=a|B=b)$ denote the conditional density of variable $A$ given $B$ at values $A=a$ and $B=b$. We note that if $A$ is discrete rather than continuous, then $f(A=a|B=b)$ is a conditional probability rather than a density. 
\begin{assumption}[Common support for $D$ and $Z_1$]\label{A2}
  \begin{align*}
  f(D=d, Z_1=z_1|M=m, X=x)>0 \quad \forall d \in \mathcal{D}\textrm, z_1 \in \mathcal{Z}_1, m \in \mathcal{M}, \textrm{ and } x \in \mathcal{X}
  \end{align*}
\end{assumption}
\begin{assumption}[Common support for $M$ and $Z_2$]\label{A3}
  \begin{align*}
    f(M=m, Z_2=z_2|D=d, X=x)>0 \quad \forall d \in \mathcal{D}, m \in \mathcal{M}, z_2 \in \mathcal{Z}_2,  \textrm{ and } x \in \mathcal{X}
  \end{align*}
\end{assumption}
\noindent Assumption \ref{A2} requires that, conditional on $M$ and $X$, observations exist with any values of $D$ and $Z_1$ that occur in the total population. Similarly, Assumption \ref{A3}  implies that, conditional on $D$ and $X$, observations exist with any values of $M$ and $Z_2$ that occur in the total population. In case of a violation of these common support conditions, effects can only be evaluated and/or tested over a subset of the support of $X$, $D$, and/or $M$. 

The next assumptions ensure that the instruments are relevant. Specifically, the treatment instrument $Z_1$ is statistically associated with the treatment $D$ conditional on covariates $X$, and the mediator instrument $Z_2$ is associated with the mediator $M$, conditional on the treatment $D$ and covariates $X$.
\begin{assumption}[Conditional dependence between the treatment and treatment instrument]\label{A4}
  \begin{align*}
  D \not\!\perp\!\!\!\perp Z_1|X=x \quad \forall  x \in \mathcal{X}
  \end{align*}
\end{assumption}
\begin{assumption}[Conditional dependence between the mediator and mediator instrument]\label{A5}
  \begin{align*}
  M \not\!\perp\!\!\!\perp Z_2|D=d, X=x  \quad \forall d \in \mathcal{D} \textrm{ and }  x \in \mathcal{X}
  \end{align*}
\end{assumption}
\noindent 
Throughout the paper, $A {\perp\!\!\!\perp} B \mid C$ denotes \emph{statistical independence} of $A$ and $B$ conditional on $C$ (i.e., a property of the joint distribution of observed data), while $A \not\!\perp\!\!\!\perp B \mid C$ denotes \emph{statistical dependence}. These symbols refer exclusively to distributional (observable) properties and do not themselves assert a causal relationship. Under the causal faithfulness postulated in Assumption~\ref{A1}, however, statistical dependence $A \not\!\perp\!\!\!\perp B \mid C$ is equivalent to $A$ and $B$ being d-connected given $C$ in the causal graph, and statistical independence is equivalent to d-separation.
Assumptions \ref{A4} and \ref{A5} can be directly tested in the data by investigating the conditional associations of $D$ and $Z_1$ as well as $M$ and $Z_2$. If these assumptions do not hold, the testing approach lacks the power to detect violations of the identifying assumptions, as discussed in more detail in \citet{huberkueck2022}.

The next assumptions invoke the conditional independence of the treatment on the one hand and the potential outcomes and potential mediators on the other hand conditional on the covariates, with ${\perp\!\!\!\perp}$ denoting statistical independence. These types of assumptions are referred to as treatment exogeneity, selection-on-observables, or unconfoundedness, as discussed in \citet{Im04} and \citet{ImWo08}. They imply that, given the covariates, the treatment is effectively random when assessing its effect on the outcome or the mediator. 
\setcounter{assumption}{5}
\begin{assumption}[Conditional independence of the treatment]\label{A6}
\begin{align*}
  \textrm{(a)} \quad & Y(d,m) {\perp\!\!\!\perp} D | X=x \quad \forall d \in \mathcal{D}, m \in \mathcal{M}, \textrm{ and } x \in \mathcal{X} \\
  \textrm{(b)} \quad & M(d) {\perp\!\!\!\perp} D | X=x \quad \forall d \in \mathcal{D}, \textrm{ and } x \in \mathcal{X}
\end{align*}
\end{assumption}
\noindent Assumption~\ref{A6} requires that conditional on covariates $X$, there exist no confounders jointly affecting the treatment $D$ on the one hand and the outcome $Y$ or the mediator $M$ on the other hand. This permits identifying the causal effect of $D$ on $M$ or $Y$ when controlling for $X$. 

Next, we impose a related conditional independence assumption also w.r.t.\ to the mediator, requiring that the latter is conditionally independent of the potential outcomes conditional on the treatment and the pre-treatment covariates. 
\begin{assumption}[Conditional independence of the mediator]\label{A7}
  \begin{align*}
  Y(d',m) {\perp\!\!\!\perp} M | D=d, X=x \quad \forall d,d' \in \mathcal{D}, m \in \mathcal{M}, \textrm{ and } x \in \mathcal{X}
  \end{align*}
\end{assumption}
\noindent Assumption \ref{A7} requires that conditional on $D$ and $X$, there exist no confounders jointly affecting the mediator $M$ and the outcome $Y$. This permits identifying the causal effect of $M$ on $Y$ when controlling for $D$ and $X$.

The next assumptions concern the validity of the first instrument $Z_1$ and require that the latter is conditionally independent of the potential outcomes and the potential mediators given the covariates $X$.

\begin{assumption}[Conditional independence of the treatment instrument]\label{A8}
\begin{align*}
  \textrm{(a)} \quad & Y(d,m) {\perp\!\!\!\perp} Z_1 | X=x \quad \forall d \in \mathcal{D}, m \in \mathcal{M}, \textrm{ and } x \in \mathcal{X} \\
  \textrm{(b)} \quad & M(d) {\perp\!\!\!\perp} Z_1 | X=x \quad \forall d \in \mathcal{D} \textrm{ and } x \in \mathcal{X}
\end{align*}
\end{assumption}
\noindent Assumption~\ref{A8} rules out confounders jointly affecting $Z_1$ on the one hand and the outcome $Y$ or the mediator $M$ on the other hand when controlling for $X$, which is similar to Assumption~\ref{A6}, which imposed such an exogeneity assumption w.r.t.\ the treatment. In addition, Assumption~\ref{A8} requires that conditional on $X$, instrument $Z_1$ does not directly affect $M$ or $Y$ other than through $D$. This exclusion restriction implies that $M(d,z_1)=M(d)$ and  $Y(d,m,z_1)=Y(d,m)$ for any value $z_1$ of instrument $Z_1$, otherwise the respective conditional independence assumptions are violated. Assumption~\ref{A8} is not sufficient for the nonparametric identification of the causal effects of $D$ on $Y$ or $M$ based on the instrument $Z_1$, which would require further assumptions like treatment monotonicity as discussed in \cite*{Imbens+94} and \cite*{Angrist+96}. However, Assumption~\ref{A8} is useful for testing the identification of the causal effects of $D$ on $M$ and $Y$ based on a selection-on-observables strategy as outlined further below. 

Lastly, we also invoke an instrument validity assumption on the second instrument $Z_2$, requiring it to be conditionally independent of the potential outcomes conditional on $D$ and $X$. 
\begin{assumption}[Conditional independence of the mediator instrument]\label{A9}
  \begin{align*}
  Y(d',m) {\perp\!\!\!\perp} Z_2 | D=d,  X=x \quad \forall d',d \in \mathcal{D}, m \in \mathcal{M}, \textrm{ and } x \in \mathcal{X}
  \end{align*} 
\end{assumption}
\noindent Assumption \ref{A9} rules out confounders that jointly affect $Z_2$ and $Y$ conditional on $D$ and $X$. It also requires that $Z_2$ does not directly influence $Y$ (other than through $M$), ensuring that the exclusion restriction $Y(d,m,z_2) = Y(d,m)$ holds for any value $z_2$ of $Z_2$. This assumption is useful for testing the identification of the causal effects of $M$ on $Y$. We note that the IV assumptions \ref{A8} and \ref{A9} are weaker than those proposed in \cite*{Miquel02}, \cite*{FroelichHuber2014}, and \cite*{rudolph2024using}, which additionally require monotonicity conditions for instrument-based identification of dynamic treatment effects or natural and interventional direct and indirect effects, respectively. By contrast, we use instruments solely for testing purposes.

If one is willing to impose the causal structure and faithfulness postulated in Assumption~\ref{A1} along with the (testable) conditional dependencies of the respective instruments and the treatment or the mediator as invoked in Assumptions \ref{A4} and Assumption \ref{A5}, then the joint satisfaction of the conditional independence assumptions on the treatment, mediators, and instruments imposed in 
Assumptions~\ref{A6}, \ref{A7}, \ref{A8}, and \ref{A9} imply the following testable implications, which under the common support assumptions \ref{A2} and \ref{A3} are verifiable for all values in the support of the respective conditioning set:
\begin{align} 
  Y &{\perp\!\!\!\perp} Z_1 | D=d, X=x \quad &\forall d \in \mathcal{D} \textrm{ and } x \in \mathcal{X},\tag{TIa}\label{TIa} \\
  M &{\perp\!\!\!\perp} Z_1 | D=d, X=x \quad &\forall d \in \mathcal{D} \textrm{ and } x \in \mathcal{X},\tag{TIb}\label{TIb} \\
   Y &{\perp\!\!\!\perp} Z_2 | D=d, M=m, X=x \quad &\forall d \in \mathcal{D}, m \in \mathcal{M} \textrm{ and } x \in \mathcal{X}. \tag{TIc}\label{TIc}
\end{align}
\noindent Furthermore, conditional on Assumptions \ref{A1}, \ref{A4}, \ref{A5}, the joint satisfaction of the testable implications (\ref{TIa}), (\ref{TIb}), (\ref{TIc}) is not only implied by but also imply the joint satisfaction of the conditional independence assumptions \ref{A6} -- \ref{A9}, as formalized in the following theorem, which is proven in the appendix.

\begin{theorem}\label{mainsetup}
$$\text{Under Assumptions }   \ref{A1}, \ref{A4}, \ref{A5}: \  \text{Assumptions } \ref{A6}, \ref{A7}, \ref{A8}, \ref{A9} \iff (\ref{TIa}),(\ref{TIb}),(\ref{TIc}).\notag $$
\end{theorem}


A subtle point concerns the interpretation of Theorem~\ref{mainsetup}, which states an equivalence between conditional independence Assumptions~\ref{A6}--\ref{A9}, formulated in terms of potential outcomes, and the testable implications (\ref{TIa})--(\ref{TIc}), which involve only factual quantities. Whether this equivalence holds in full depends on the underlying counterfactual framework.

Under a nonparametric structural equation model with independent errors (NPSEM-IE) in the sense of \citet{Pearl00}, every variable is generated by a structural function of its parents and a mutually independent exogenous error. The structural model then determines the joint distribution of all potential outcomes, including \emph{cross-world} quantities such as $Y(d,m)$ and $M(d')$ for $d\neq d'$. In this case, the d-separation relations encoded in Assumption~\ref{A1} together with faithfulness propagate to all counterfactual independencies, and the equivalence stated in Theorem~\ref{mainsetup} is exact. We also stress that as long as we are interested in estimands that involve cross-world quantities, such as the natural direct or indirect effects in the mediation analysis, some form of cross-world independence is necessary for the identification.\footnote{Some researchers argue against such estimands \citep{RobinsRichardson2010, robins2022interventionist}. We refer the reader to \cite{andrews2021insights} for a more detailed discussion of the cross-world independence assumption in causal mediation analysis.}

NPSEM-IE is, however, a strong assumption: the mutual independence of structural errors implicitly imposes independence across counterfactual worlds. If one only adopts the weaker FFRCISTG framework of \citet{Robins86} and \citet{RobinsRichardson2010}, in which potential outcomes are assumed to exist only under single-world interventions and joint independence across worlds is not imposed, then Theorem~\ref{mainsetup} should be read as a statement about \emph{factual} potential outcomes, i.e., $Y(d,m)$ and $M(d)$ among subjects actually receiving $D=d$ and $M=m$. In this case, our test verifies necessary but not sufficient conditions for Assumptions~\ref{A6}--\ref{A9}: violations of conditional independence that exclusively concern counterfactual values $d'\neq d$ or $m'\neq m$ for subjects with factual $D=d$ and $M=m$ cannot be detected.

From a practical perspective, however, it seems unlikely that violations exclusively occur among counterfactual, but not among factual, outcomes and mediators, because this would imply rather implausible modeling constraints. We would therefore expect our test to have power in empirical applications, as violations of conditional independence should typically not affect only counterfactual outcomes or mediators, but also at least some factual ones.

Beyond the joint equivalence stated in Theorem~\ref{mainsetup}, the proof structure yields a useful by-product which permits localizing the rejection in terms of which identifying assumption is violated, summarized in the following corollary.

\begin{corollary}[Diagnostic decomposition]\label{cor:decomposition}
Under Assumption~\ref{A1}:\vspace{-1.5em}
\begin{align*}
\text{Under Assumption~\ref{A4}:} \quad & (\ref{TIa}) \implies \ref{A6}(a) \text{ and } \ref{A8}(a), \\
\text{Under Assumption~\ref{A4}:} \quad & (\ref{TIb}) \implies \ref{A6}(b) \text{ and } \ref{A8}(b), \\
\text{Under Assumption~\ref{A5}:} \quad & (\ref{TIc}) \implies \ref{A7} \text{ and } \ref{A9}.
\end{align*}
\end{corollary}
\vspace{-1em}
\noindent Corollary~\ref{cor:decomposition} follows directly from the proof of Theorem~\ref{mainsetup}, where each implication is established separately. (\ref{TIa}) flags the treatment-outcome arm of the model, (\ref{TIb}) flags the treatment-mediator arm, and (\ref{TIc}) flags the mediator-outcome arm. The decomposition is one-directional, because the converse direction in Theorem~\ref{mainsetup} requires the joint assumption set.

We note that if the conditional independence assumptions \ref{A6} - \ref{A9} 
hold for both factual and counterfactual mediators or outcomes, the following causal effects on the distributions of the mediator or the outcome are identified:
\begin{itemize}
	\item The effect of treatment $D$ on outcome $Y$, e.g., the ATE of treatment levels $d \neq d'$, $E[Y(d,M(d)) - Y(d',M(d'))]$, where identification follows from the conditional independence of the treatment invoked in Assumption~\ref{A6}(a).
	\item The effect of treatment $D$ on mediator $M$, e.g., the ATE $E[M(d) - M(d')]$, where identification follows from the conditional independence of the treatment invoked in Assumption~\ref{A6}(b).
		\item The effect of mediator $M$ on outcome $Y$, e.g., the ATE of mediator levels $m \neq m'$, $E[Y(D,m) - Y(D,m')]$, where identification follows from the conditional independence of the mediator invoked in Assumption \ref{A7}.
		\item The dynamic effect of specific sequences of treatment $D$ and mediator $M$ on outcome $Y$, e.g., the ATE $E[Y(d,m) - Y(d',m')]$, including the average controlled direct effect of the treatment on the outcome conditional on fixing the mediator at some value $M=m$, $E[Y(d,m) - Y(d',m)]$, where identification follows from Assumptions~\ref{A6}(a) and \ref{A7}, see for instance the discussions in \citet{LechnerMiquel2010} and \citet{huber2023causal}, Chapters 4.9 and 4.10.
	   \item The effect of treatment $D$ on outcome $Y$ in sample selection models, where $M$ represents a binary indicator for the observability of $Y$ (but not a mediator, because $M$ does not affect $Y$ in sample selection models). Identification follows from Assumptions~\ref{A6}(a) and \ref{A7} when replacing $Y(d,m)$ by $Y(d)$ (as $M$ does not affect $Y$ in sample selection models), see for instance the discussions in \citet{bia2023double} and \citet{huber2023causal}, Chapter 4.11. 
\end{itemize}
In causal mediation analysis, the controlled direct effect, which is based on forcing or prescribing the mediator to take a specific value $M=m$, is typically not the only causal parameter of interest. A large literature also focuses on natural direct and indirect effects defined in terms of potential (rather than prescribed) mediator states, for instance, discussed in \cite{RoGr92}, \cite{Pearl01}, \citet{ImKeYa10}, and \citet{Huber2012}. For instance, the average natural direct treatment effect of $d\neq d'$ on $Y$ conditional the potential mediator under treatment value $d$, $M(d)$, corresponds to $E[Y(d,M(d)) - Y(d',M(d))]$, while the average natural indirect effect of $D$ on $Y$ via $M$ when fixing the treatment at $D=d'$ is  $E[Y(d,M(d)) - Y(d,M(d'))]$. 

It is worth noting that the identification of natural effects not only hinges on the satisfaction of Assumptions \ref{A6} and \ref{A7} for both factual and counterfactual outcomes, but even on additional requirements. One additional condition that yields identification and has been suggested in  \cite{Pearl01} is invoking conditional independence of potential mediators and potential outcomes across treatment states $d\neq d'$:
\begin{align}\label{crossworld}
	Y(d,m)& {\perp\!\!\!\perp} M(d') | X=x  \quad \forall d,d' \in \mathcal{D}, m \in \mathcal{M}, \textrm{ and } x \in \mathcal{X}.
\end{align}
This conditional independence is an inherently counterfactual assumption, as either only $d$ or $d'$ can be observed for any subject in the sample. For this reason, we cannot test this assumption in the data. Furthermore, it is worth noting that the conditional independence in \eqref{crossworld} as well as Assumption~\ref{A6} is implied by the following, joint conditional independence assumption of \citet{ImKeYa10}, which they impose in addition to Assumption \ref{A7} for the identification of natural direct and indirect effects:
\begin{align}\label{joint}
\{Y(d,m), M(d')\} & {\perp\!\!\!\perp} D | X=x  \quad \forall d,d' \in \mathcal{D}, m \in \mathcal{M}, \textrm{ and } x \in \mathcal{X}.
\end{align}
The conditional independence assumption in expression \eqref{joint} implies that the joint distribution of potential outcomes and potential mediators is conditionally independent of treatment assignment. In contrast, Assumption~\ref{A6} only imposes conditional independence w.r.t.\ the marginal distributions of the potential outcomes and potential mediators. Yet, we argue that testing Assumption~\ref{A6} for actual outcomes typically also has nontrivial power against violations of the counterfactual condition \eqref{joint}. The reason is that only under quite specific mediator and outcome models, it can be the case that Assumption~\ref{A6} always holds for factual mediators and outcomes, while we have at the same time that the conditional independence \eqref{joint} is violated for certain counterfactual outcomes and mediators. For instance, if the treatment is randomly assigned given $X$, then both the joint conditional independence assumption \eqref{joint} as well as Assumption~\ref{A6} on the marginal distributions are satisfied. However, if conditional randomization of $D$ fails, then we would suspect violations of all assumptions \eqref{joint} and \ref{A6}, and even so for at least some factual outcomes and mediators.

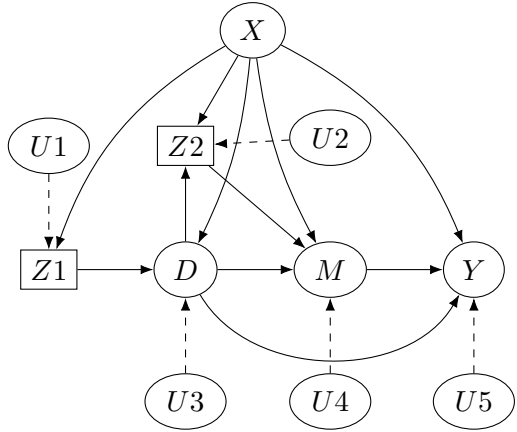
\begin{figure}[!htp]
	\centering
	\caption{Causal models satisfying the identification results in Theorem \ref{mainsetup}}
	\label{dag_1}
		\centering
		\begin{tikzpicture}
			\node[state,rectangle] (Z1) at (0,0) {$Z1$};
			\node[state] (D) [right =of Z1] {$D$};
			\node[state] (M) [right =of D] {$M$};
			\node[state] (Y) [right =of M] {$Y$};
			\node[state,rectangle] (Z2) [above =of D] {$Z2$};
			\node[state] (X) [above right =of Z2,xshift=-0.8cm] {$X$};
			\node[state] (U1) [above =of Z1] {$U1$};
			\node[state] (U2) [above =of M] {$U2$};
			\node[state] (U4) [below =of M] {$U4$};
			\node[state] (U5) [below =of Y] {$U5$};
			\node[state] (U3) [below =of D] {$U3$};
			\path (Z1) edge (D);
			\path (D) edge (M);
			\path (M) edge (Y);
			\path (X) edge (Z2);
			\path (X)[bend right=20] edge (Z1);
			\path (X)[bend left=10] edge (D);
			\path (X)[bend right=10] edge (M);
			\path (X)[bend left=20] edge (Y);
			\path[dashed] (U1) edge (Z1);
			\path[dashed] (U4) edge (M);
			\path[dashed] (U5) edge (Y);
			\path[dashed] (U3) edge (D);
			\path (Z2) edge (M);
			\path (D) edge (Z2);
			\path (D) edge[bend right=60] (Y);
			\path[dashed] (U2) edge (Z2);
		\end{tikzpicture}
\end{figure}

Figure~\ref{dag_1} presents a causal model in which the identifying assumptions of the various causal effects discussed above as well as the testable implications given in Theorem \ref{mainsetup} are satisfied, based on a causal graph in which causal relations between variables are represented by arrows, see e.g.\ \cite{Pearl00}. Treatment $D$ affects outcome $Y$ both directly and via the mediator $M$ and is conditionally independent of potential mediators and outcomes because no unobserved variables, whose effects are depicted by dashed lines (to indicate their non-observability), jointly affect $D$ and the post-treatment variables $M$ and $Y$ conditional on covariates $X$. Analogously, there are no unobservables jointly affecting the mediator $M$ and the outcome $Y$ conditional on $X$ and $D$. Furthermore, there are no confounders of the instrument for the treatment, $Z_1$, and the $M$ or $Y$ conditional on $X$, and $Z_1$ does not directly affect $M$ or $Y$ other than through $D$.  Moreover, there are no confounders of the instrument for the mediator, $Z_2$, and $Y$ conditional on $X$ and $D$, and $Z_2$ does not directly affect $Y$ other than through $M$. 

We wish to stress that our assumptions do not pin down a unique causal structure but rather restrict the space of admissible causal structures. Each assumption eliminates certain causal structures, and it is their joint satisfaction that narrows this space sufficiently for identification and testing.


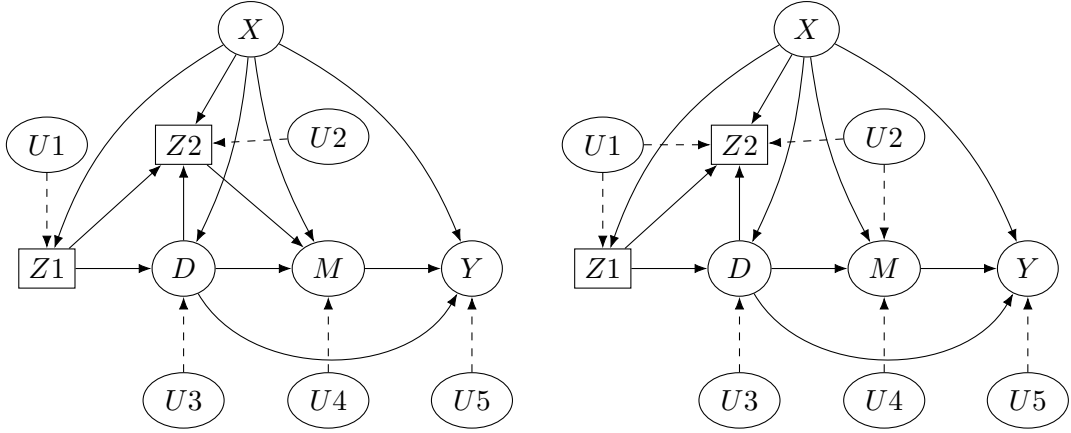
\begin{figure}[!htp]
	\centering \caption{\label{dag_2}  Causal models with conditional associations between the instruments \bigskip}
\begin{subfigure}
	\centering
\begin{tikzpicture}
  \node[state,rectangle] (Z1) at (0,0) {$Z1$};

  \node[state] (D) [right =of Z1] {$D$};
  \node[state] (M) [right =of D] {$M$};
  \node[state] (Y) [right =of M] {$Y$};
  \node[state,rectangle] (Z2) [above =of D] {$Z2$};
  \node[state] (X) [above right =of Z2,xshift=-0.8cm] {$X$};
  \node[state] (U1) [above =of Z1] {$U1$};
    \node[state] (U2) [above =of M] {$U2$};
  \node[state] (U4) [below =of M] {$U4$};
  \node[state] (U5) [below =of Y] {$U5$};
\node[state] (U3) [below =of D] {$U3$};
   \path (Z1) edge (D);
   \path (D) edge (M);
   \path (M) edge (Y);
   \path (X) edge (Z2);
   \path (Z1) edge (Z2);
   \path (X)[bend right=20] edge (Z1);
   \path (X)[bend left=10] edge (D);
   \path (X)[bend right=10] edge (M);
   \path (X)[bend left=20] edge (Y);
   \path[dashed] (U1) edge (Z1);
   \path[dashed] (U4) edge (M);
   \path[dashed] (U5) edge (Y);
    \path[dashed] (U2) edge (Z2);
   \path (Z2) edge (M);
   \path (D) edge (Z2);
   \path (D) edge[bend right=60] (Y);
   \path[dashed] (U3) edge (D);
\end{tikzpicture} 
	\end{subfigure}	
\quad
\begin{subfigure}
	\centering
	\begin{tikzpicture}
		\node[state,rectangle] (Z1) at (0,0) {$Z1$};
		
		\node[state] (D) [right =of Z1] {$D$};
		\node[state] (M) [right =of D] {$M$};
		\node[state] (Y) [right =of M] {$Y$};
		\node[state,rectangle] (Z2) [above =of D] {$Z2$};
		\node[state] (X) [above right =of Z2,xshift=-0.8cm] {$X$};
		\node[state] (U1) [above =of Z1] {$U1$};
		\node[state] (U2) [above =of M] {$U2$};
		 \node[state] (U4) [below =of M] {$U4$};
		\node[state] (U3) [below =of D] {$U3$};
		 \node[state] (U5) [below =of Y] {$U5$};
		\path (Z1) edge (D);
		\path (D) edge (M);
		\path (M) edge (Y);
		\path (X) edge (Z2);
		\path (Z1) edge (Z2);
		\path (X)[bend right=20] edge (Z1);
		\path (X)[bend left=10] edge (D);
		\path (X)[bend right=10] edge (M);
		\path (X)[bend left=20] edge (Y);
		\path[dashed] (U1) edge (Z1);
		\path[dashed] (U1) edge (Z2);
		\path[dashed] (U2) edge (Z2);
		\path[dashed] (U2) edge (M);
		\path (D) edge (Z2);
		\path (D) edge[bend right=60] (Y);
		\path[dashed] (U3) edge (D);
		\path[dashed] (U4) edge (M);
		\path[dashed] (U5) edge (Y);
	\end{tikzpicture} 
	\end{subfigure}	
	
\end{figure}

It is worth noting that if $Z_2$ has a direct causal effect on $M$ as in Figure~\ref{dag_1}, then the satisfaction of the identifying assumptions underlying Theorem \ref{mainsetup} requires that $Z_1$ and $Z_2$ are conditionally independent of each other given $D$ and $X$. To see this, note that if e.g.\ $Z_1$ has a direct effect on $Z_2$ conditional on $D$ and $X$ and $Z_2$ has a direct effect on $M$, $Z_1$ is a confounder that jointly affects treatment $D$ (by Assumption~\ref{A4}) on the one hand and $M$ and $Y$ on the other hand. Related issues also occur if unobserved confounders affect both $Z_1$ and $Z_2$ and $Z_2$ directly affects $M$. The left graph of Figure~\ref{dag_2} provides such a causal model in which Assumption~\ref{A6} is violated. 

If, on the contrary, $Z_2$ is associated with $M$ solely through unobserved confounders of both $Z_2$ and $M$ (rather than through a direct effect), then all identifying assumptions in Theorem \ref{mainsetup} hold even if $Z_1$  has a direct causal effect on $Z_2$ given $D$ and $X$, and/or unobserved confounders affect both $Z_1$ and $Z_2$, as illustrated in the right graph of Figure~\ref{dag_2}. 
The rationale behind this is that $Z_2$ does not lie on any causal pathway through which the treatment $D$ influences the mediator $M$, as $Z_2$ does not affect $M$. Here, adjusting for $Z_2$ would even be harmful as it would introduce so-called M-bias, a specific form of collider or selection bias, see e.g. \cite{Pearl00}. The reason is that by conditioning on $Z_2$, one introduces a spurious association between the treatment $D$ and the unobservable $U_2$ which also affects $M$ (and $Y$ via $M$).  Such a spurious association comes from the fact that $D$ either directly affects the collider variable $Z_2$ or is associated with the latter through $U_1$ (which both affects $Z_2$ and $D$ via $Z_1$).


\section{Modifications when controlling for the mediator instrument}\label{Assumptions2}


In causal models where the first instrument $Z_1$ affects the second instrument $Z_2$, and $Z_2$, in turn, affects the mediator $M$, it is necessary to control for $Z_2$ to block any causal effects of $Z_1$ on $M$ or $Y$ that do not operate via the treatment $D$, conditional on covariates $X$. This scenario is illustrated in the left graph of Figure~\ref{dag_2}, where controlling for $Z_2$ does not introduce collider bias, in contrast to the scenario depicted in the right graph of  Figure \ref{dag_2}. Consequently, an alternative testing approach for identification may involve including $Z_2$ in certain conditioning sets when verifying the conditional independence of $Z_1$ and $Y$. 
Accordingly, we modify the conditional independence assumptions for the treatment and the first instrument to hold when controlling for both $X$ and $Z_2$, rather than $X$ alone. To achieve this, we replace Assumptions~\ref{A6} and~\ref{A8} with the following modified assumptions:


\setcounter{assumption}{6}
\begin{assumptionmod}[Conditional independence of the treatment (modified)]\label{A6m}
\begin{align*}
  \textrm{(a)} \quad & Y(d,m) {\perp\!\!\!\perp} D | Z_2=z_2, X=x \quad \forall d \in \mathcal{D}, m \in \mathcal{M}, z_2 \in \mathcal{Z}_2, \textrm{ and } x \in \mathcal{X} \\
  \textrm{(b)} \quad & M(d) {\perp\!\!\!\perp} D | Z_2=z_2, X=x \quad \forall d \in \mathcal{D}, z_2 \in \mathcal{Z}_2, \textrm{ and } x \in \mathcal{X}
\end{align*}
\end{assumptionmod}
\setcounter{assumption}{8}
\begin{assumptionmod}[Conditional independence of the treatment instrument (modified)]\label{A8m}
\begin{align*}
  \textrm{(a)} \quad & Y(d,m) {\perp\!\!\!\perp} Z_1 | Z_2=z_2, X=x \quad \forall d \in \mathcal{D}, m \in \mathcal{M}, z_2 \in \mathcal{Z}_2, \textrm{ and } x \in \mathcal{X} \\
  \textrm{(b)} \quad & M(d) {\perp\!\!\!\perp} Z_1 | Z_2=z_2, X=x \quad \forall d \in \mathcal{D}, z_2 \in \mathcal{Z}_2, \textrm{ and } x \in \mathcal{X}
\end{align*}
\end{assumptionmod}
These adjusted identifying assumptions lead to the following testable implications, which replace the previous implications \eqref{TIa} and \eqref{TIb}. The earlier implication \eqref{TIc} remains unchanged:
\begin{align}
	Y {\perp\!\!\!\perp} Z_1 | D=d, Z_2=z_2, X=x  \quad \forall d \in \mathcal{D}, z_2 \in \mathcal{Z}_2, \textrm{ and }  x \in \mathcal{X} , \tag{TIam} \label{TIam} \\
	M {\perp\!\!\!\perp} Z_1 | D=d, Z_2=z_2, X=x  \quad \forall d \in \mathcal{D}, z_2 \in \mathcal{Z}_2, \textrm{ and }  x \in \mathcal{X} . \tag{TIbm} \label{TIbm}
\end{align}
Additionally, there is another testable implication, which follows from the principle that any causal association between $Z_1$ and $Y$ via $Z_2$ must operate through $M$: 
\begin{align}
	Y {\perp\!\!\!\perp} Z_1 | D=d, M=m, X=x \quad \forall d \in \mathcal{D},  m \in \mathcal{M}, \textrm{ and } x \in \mathcal{X}. \tag{TId} \label{TId}
\end{align}
However, the testable implication (\ref{TId}) does not provide additional information or enhance testing power compared to the earlier testable implications. This is formalized in the following lemma, which states that given  Assumption \ref{A1} and implications (\ref{TIam}), (\ref{TIbm}), (\ref{TIc}) implies (\ref{TId}) and vice versa:
\begin{lemma}\label{lemma1}
	\begin{align}
		\textrm{Given the causal structure outlined in Assumption } \ref{A1}: (\ref{TIam}),(\ref{TIbm}),(\ref{TIc}) \implies (\ref{TId}) \notag \\
		\textrm{Given the causal structure outlined in Assumption } \ref{A1}:(\ref{TIbm}),(\ref{TIc}),(\ref{TId}) \implies (\ref{TIam}) \notag
	\end{align}
\end{lemma}
The following theorem formalizes that our modified set of identifying assumptions, which involve conditioning on $Z_2$, implies and is implied by the respective testable implications:
\begin{theorem}\label{altsetup}
\begin{align}
\vspace{-1.5em}
\text{Under Assumptions }\ref{A1}, \ref{A4}, \ref{A5}: \  \text{Assumptions }  \ref{A6m}, \ref{A7}, \ref{A8m}, \ref{A9} &\iff (\ref{TIam}),(\ref{TIbm}),(\ref{TIc}) \notag \\
 &\iff (\ref{TIbm}),(\ref{TIc}),(\ref{TId}).\notag 
  \end{align}
\end{theorem}
The proofs of Lemma \ref{lemma1} and Theorem \ref{altsetup} are provided in the appendix.

\begin{corollary}[Diagnostic decomposition, alternative setup]\label{cor:decomposition_alt}
Under Assumption~\ref{A1}:\vspace{-1.5em}
\begin{align*}
\text{Under Assumption~\ref{A4}:} \quad & (\ref{TIam}) \implies \ref{A6m}(a) \text{ and } \ref{A8m}(a), \\
\text{Under Assumption~\ref{A4}:} \quad & (\ref{TIbm}) \implies \ref{A6m}(b) \text{ and } \ref{A8m}(b), \\
\text{Under Assumption~\ref{A5}:} \quad & (\ref{TIc}) \implies \ref{A7} \text{ and } \ref{A9}.
\end{align*}
\end{corollary}
\vspace{-1em}

It is important to note that while the testable implications involve conditioning on $Z_2$, this does not necessarily mean that controlling for $Z_2$ is also appropriate for the identification of causal effects. Indeed, if the testable implications in Theorem \ref{altsetup} hold and $Z_2$ is a confounder of $D$ and $M$, or if unobserved confounders affect both $D$ and $Z_2$ while $Z_2$ affects $M$, then controlling for $Z_2$ (along with $X$) is necessary when evaluating the causal effect of $D$ on $M$ or $Y$. However, if $D$ affects $Z_2$ and there are no unobserved confounders between $D$ and $Z_2$, then controlling for $Z_2$ is necessary only for testing, but not for evaluating the causal effects of $D$. In such cases, conditioning on $Z_2$ would block any effects of $D$ on $M$ and $Y$ that operate through $Z_2$, thereby limiting the identification to a partial effect.

\section{Pre- and post-treatment covariates}\label{Assumptions3}

\begin{figure}[!htp]
	\centering \caption{\label{dag_5}  Pre- and post-treatment covariates \bigskip}
\centering 
\begin{tikzpicture}
  \node[state,rectangle] (Z1) at (0,0) {$Z1$};

  \node[state] (D) [right =of Z1] {$D$};
  \node[state] (M) [right =of D] {$M$};
  \node[state] (Y) [right =of M] {$Y$};
  \node[state, rounded corners] (W) [above =of D] {$W$};
  \node[state,rectangle] (Z2) [above =of M] {$Z2$};
  \node[state] (U1) [above =of Z1] {$U1$};
  \node[state, rounded corners] (X) [above right=of W,xshift=-0.8cm] {$X$};
     	\node[state] (U4) [below =of M] {$U4$};
  \node[state] (U5) [below =of Y] {$U5$};
  \node[state] (U3) [below =of D] {$U3$};
  \node[state] (U2) [above =of Y] {$U2$};

   \path (Z1) edge (D);
   \path (D) edge (M);
   \path (M) edge (Y);
   \path (X) edge (Z2);
   \path (X)[bend right=20] edge (Z1);
   \path (X)[bend left=10] edge (D);
   \path (X)[bend right=10] edge (M);
   \path (X)[bend left=20] edge (Y);
   \path[dashed] (U1) edge (Z1);
   \path[dashed] (U4) edge (M);
   \path[dashed] (U5) edge (Y);
      \path[dashed] (U3) edge (D);
      \path[dashed] (U2) edge (Z2);
   \path (W) edge (Z2);
   \path (W) edge (M);
   \path (W) edge (Y);
   \path (D) edge (W);
   \path (X) edge (W);
   \path (Z2) edge (M);
   \path (D) edge (Z2);
   \path (Z1) edge (W);
   \path (D) edge[bend right=60] (Y);
\end{tikzpicture} 
\end{figure}
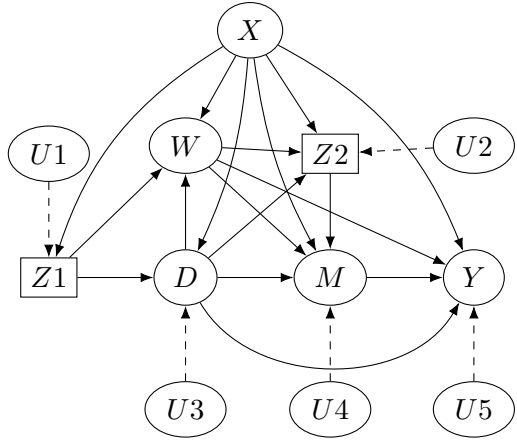

In many empirical applications, pre-treatment covariates may be insufficiently informative to fully control for confounders jointly affecting the mediator $M$ and the outcome $Y$. As a result, the conditional independence assumption for the mediator (Assumption \ref{A7}) may be violated. This issue becomes particularly relevant when the mediator is measured at a substantially later time than the treatment. Given that control variables for the treatment are typically measured shortly before treatment assignment, it is similarly reasonable to control for potential confounders of the mediator-outcome relationship shortly before selection into the mediator. In such scenarios, several control variables for the mediator may be influenced by the treatment. For instance, when assessing the treatment effect of education ($D$) on earnings ($Y$) with work experience as a mediator ($M$), post-treatment covariates like health or mental well-being ($W$) might affect both $M$ and $Y$. For this reason, the subsequent discussion considers the case that the treatment can influence observed post-treatment confounders of the mediator-outcome relation to be controlled for, which are denoted by $W$, while $\mathcal{W}$ denotes their support. Figure \ref{dag_5} presents a causal graph illustrating both pre-treatment covariates $X$ and post-treatment covariates $W$. The covariate (set) $W$ may be affected by treatment $D$, pre-treatment covariates $X$, and the first instrument $Z_1$, and in turn, may affect the mediator $M$, outcome $Y$, and the second instrument $Z_2$. 

In this testing approach, post-treatment covariates $W$ must now be included in the conditioning set when imposing the conditional independence assumptions for the second instrument and the mediator. For the test to have power, there must be a first-stage association between instrument $Z_2$ and treatment $M$, even when controlling for post-treatment covariates $W$ in addition to the treatment $D$ and pre-treatment covariates $X$. We also rule out any direct effect of $M$ on $W$, given that $W$ consists of pre-mediator confounders and thus cannot be causally influenced by the mediator. Given the limitations of pre-treatment confounders discussed above, we now introduce modified assumptions that account for post-treatment variables. Specifically, we replace Assumptions~\eqref{A1}, \eqref{A5}, \eqref{A7}, and \eqref{A9} in Section~\ref{Assumptions1} with the following four assumptions: 

\setcounter{assumption}{1}
\begin{assumptionmod}[Causal structure and faithfulness]\label{A1m}
\begin{align*}
  M(y)=M, D(m, y, z_2)=D, X(d, m, y, z_2)=X,  Z_1(d, m, y, z_2)=Z_1,   Z_2(m, y)=Z_2,  \\ 
  \; \textrm{ and } W(m) = W,\\
  \forall d \in \mathcal{D}, m \in \mathcal{M},   z_2 \in \mathcal{Z}_2  \textrm{ and }  y  \in \mathcal{Y},\\
\end{align*}
  \\[-2cm]only variables which are d-separated in any causal model are statistically independent.
\end{assumptionmod}

\setcounter{assumption}{5}
\begin{assumptionmod}[Conditional dependence between the mediator and mediator instrument]\label{A5m} 
	\begin{align*}
		M \not\!\perp\!\!\!\perp Z_2|D=d, X=x, W=w  \quad \forall d \in \mathcal{D}, x \in \mathcal{X}, \textrm{ and } w \in \mathcal{W} 
	\end{align*}
\end{assumptionmod}

\setcounter{assumption}{7}
\begin{assumptionmod}[Conditional independence of the mediator]\label{A7m} 
	\begin{align*}
		Y(d',m) {\perp\!\!\!\perp} M | D=d, X=x, W=w \quad \forall d',d \in \mathcal{D}, m \in \mathcal{M}, x \in \mathcal{X}, \textrm{ and } w \in \mathcal{W} 
	\end{align*}
\end{assumptionmod}
\setcounter{assumption}{9}
\begin{assumptionmod}[Conditional independence of the mediator instrument]\label{A9m} 
	\begin{align*}
		Y(d',m) {\perp\!\!\!\perp} Z_2 | D=d,  X=x, W=w \quad \forall d',d \in \mathcal{D}, m \in \mathcal{M}, x \in \mathcal{X}, \textrm{ and } w \in \mathcal{W} 
	\end{align*}
\end{assumptionmod}
With these modifications, which include $W$ in the conditioning set, the testable implication~\eqref{TIc} is replaced by the following implication, which asserts that the second instrument is conditionally independent of the outcome given the treatment, mediator, and both pre- and post-treatment covariates: 
\begin{align*} 
	Y {\perp\!\!\!\perp} Z_2 | D=d, M=m, X=x, W=w \quad \forall d \in \mathcal{D}, m \in \mathcal{M}, x \in \mathcal{X}, \textrm{ and } w \in \mathcal{W}. \tag{TIe} \label{TIe}
\end{align*}
Similar to the previously analyzed causal scenarios, our revised set of identifying assumptions, which includes conditioning on $W$, implies and is implied by a corresponding set of testable implications. This result is formally stated in the following theorem.
\begin{theorem}\label{dynsetup}
  \begin{align}
\text{Under Assumptions }   \ref{A1m}, \ref{A4}, \ref{A5m}: \  \text{Assumptions } \ref{A6}, \ref{A7m}, \ref{A8}, \ref{A9m} &\iff (\ref{TIa}),(\ref{TIb}),(\ref{TIe}) \notag 
  \end{align}
\end{theorem}
The proof can be found in the appendix.
\begin{corollary}[Diagnostic decomposition, dynamic setup]\label{cor:decomposition_dyn}
Under Assumption~\ref{A1m}:\vspace{-1.5em}
\begin{align*}
\text{Under Assumption~\ref{A4}:} \quad & (\ref{TIa}) \implies \ref{A6}(a) \text{ and } \ref{A8}(a), \\
\text{Under Assumption~\ref{A4}:} \quad & (\ref{TIb}) \implies \ref{A6}(b) \text{ and } \ref{A8}(b), \\
\text{Under Assumption~\ref{A5m}:} \quad & (\ref{TIe}) \implies \ref{A7m} \text{ and } \ref{A9m}.
\end{align*}
\end{corollary}
\vspace{-1em}

It is important to note that if the conditional independence assumptions \ref{A6}, \ref{A7m}, \ref{A8}, and \ref{A9m} hold for both factual and counterfactual mediators or outcomes, the effect of treatment $D$ on outcome $Y$ and mediator $M$ (by Assumption~\ref{A6}), the effect of the mediator $M$ on outcome $Y$ (by Assumption \ref{A7m}), and the dynamic effect of specific sequences of $D$ and $M$, including the controlled direct effect, (by Assumptions~\ref{A6}(a) and \ref{A7m}) are identified, as well as the effect of treatment $D$ on outcome $Y$ in sample selection models, where $M$ represents a binary indicator for the observability of $Y$ (by Assumptions~\ref{A6}(a) and \ref{A7m}). See, for instance, \citet{LechnerMiquel2010}, \citet{bia2023double}, and \citet{huber2023causal} for discussions of identification in dynamic treatment and sample selection models with post-treatment control variables. Regarding natural direct and indirect effects, identification remains unattainable under post-treatment confounding, even when imposing the joint conditional independence assumption in expression \eqref{joint} along with Assumption \ref{A7m} and analogous conditional independence assumption on $W$. However, unlike the situation where all confounders are pre-treatment, identification under post-treatment confounders is infeasible without imposing specific parametric conditions. See, for instance, the discussions in \citet{AvinShpitserPearl2005}, \citet{Robins2003}, who imposes the absence of treatment-mediator interaction effects on the outcome to obtain identification, and \citet{ImYa2011}, who assume that any treatment-mediator interaction effects are homogeneous across subjects.

\section{Testing}\label{testing}

In this section, we present our testing approach, concentrating on the assumptions and testable implications detailed in Section \ref{Assumptions1}.  Nonetheless, the causal scenarios outlined in Sections \ref{Assumptions2} and \ref{Assumptions3} can also be incorporated by suitably modifying the conditioning sets during testing. Instead of testing for full statistical independence as specified in testable implications \eqref{TIa}, \eqref{TIb}, and \eqref{TIc}, we test the conditional mean independence of the outcome or mediator and the instruments, which suffices for identifying conditional or unconditional average causal effects of the treatment and the mediator.\footnote{Testing conditional independence is a challenging practical problem, especially when the conditioning set involves many continuous variables \citep{shah2020hardness}.} To formalize the testable implications in terms of conditional mean independence, let \( \mu_B(a) = E(B|A=a) \) represent the conditional mean of the random variable \( B \), where \( A \) denotes a set of conditioning variables with values \( a \). Then, testing whether implications \eqref{TIa}, \eqref{TIb}, \eqref{TIc} hold on average involves verifying the following null hypothesis.
\begin{align}\label{H0}
H_0: &&\mu_Y(d,x)-\mu_Y(d,x,z_1)=0 \quad  \forall d \in \mathcal{D}, z_1 \in \mathcal{Z}_1, \textrm{ and } x \in \mathcal{X}, \notag\\
   &&\mu_M(d,x)-\mu_M(d,x,z_1)=0  \quad \forall d \in \mathcal{D}, z_1 \in \mathcal{Z}_1 \textrm{ and } x \in \mathcal{X}, \notag\\
   &&\mu_Y(d,m,x)-\mu_Y(d,m,x,z_2)=0  \quad \forall d \in \mathcal{D},  m \in \mathcal{M}, z_2 \in \mathcal{Z}_2, \textrm{ and } x \in \mathcal{X}  . \tag{H0}
\end{align}


The intuition underlying hypothesis \eqref{H0} is that if \( Y \) and \( M \) are conditionally mean independent of the instruments, then the inclusion or exclusion of the instruments in the respective conditioning sets should not affect the conditional means of the outcome and mediator. While a rejection of the joint $H_0$ does not by itself indicate which identification assumption is violated, Corollaries~\ref{cor:decomposition}--\ref{cor:decomposition_dyn} show that the three components of $H_0$ have a diagnostic interpretation: a rejection of the first component flags the treatment-outcome arm of the model, a rejection of the second flags the treatment-mediator arm, and a rejection of the third flags the mediator-outcome arm. The components can therefore be reported separately as a localizing diagnostic in addition to the joint test.

Following \citet{huberkueck2022}, our test relies on a quadratic formulation of the null hypothesis, which is a common approach in specification tests for nonparametric regression, see e.g. \citet{Ra97}, \citet{RaHaLi06}, \citet{hong1995consistent} and \citet{wooldridge1992test}. To this end, let us denote by 
\begin{align}\label{H0_theta}
\theta=E 
\begin{pmatrix}
	{\left(\mu_Y(D,X)-\mu_Y(D,X,Z_1)\right)}^2\\
	{\left(\mu_M(D,X)-\mu_M(D,X,Z_1)\right)}^2\\
	{\left(\mu_Y(D,M,X)-\mu_Y(D,M,X,Z_2)\right)}^2
\end{pmatrix},
\end{align}
which implies the null hypothesis
\begin{align}\label{H0_quadratic}
H_0: \theta =0.
\end{align}

We note that testing based on $\theta$ can accommodate both discrete and continuous variables for the instruments, treatment, mediator, outcome, and covariates.  In contrast, \citet{huberkueck2022} focused on a binary instrument, testing the squared difference in conditional mean outcomes with instrument values of zero versus one, rather than comparing the inclusion versus exclusion of (possibly non-binary) instruments, as proposed here. In line with \citet{huberkueck2022}, we test the null hypothesis \eqref{H0_quadratic} by employing a moment condition using the following score function to determine whether it is mean zero:
\begin{align}\label{score2}
	\phi(V,\theta,\eta)={(\eta_1(V)-\eta_2(V))}^2-\theta+\zeta,
\end{align}
where $V=(Y,D,Z_1,M,Z_2,X)$ represents the random variables, while $\eta_1(V)=(\mu_Y(D,X)$,   $\mu_M(D,X)$, $\mu_Y(D,M,X))'$, $\eta_2(V)=(\mu_Y(D,X,Z_1),\mu_M(D,X,Z_1), \mu_Y(D,M,X,Z_2))'$ are the respective conditional means. Additionally, $\zeta$ is an independent mean-zero random variable with $\|\zeta\|_{P,q}<C$, where $C$ is a positive constant, for $q>2$ and variance $\sigma_\zeta^2>0$.\footnote{For any random vector $R = (R_1,...,R_l),$ we have $||R||_q = \max_{1\leq j \leq l}||R_l||_q, $ where $||R_l||_q = (E[|R_l|^q])^{\frac{1}{q}}.$}

The random perturbation \( \zeta \) is introduced to avoid a degenerate variance in our testing approach under the null hypothesis \( H_0 \). The variance \( \sigma_\zeta^2 \) acts as a tuning parameter that should be adjusted based on the sample size \( n \). This adjustment involves a trade-off: a higher variance brings the testing based on \( \phi \) closer to the nominal size in smaller samples, but it may diminish the test's power by potentially obscuring genuine deviations from the null hypothesis. As shown in \citet{huberkueck2022}, score functions of this quadratic type satisfy Neyman orthogonality under the null hypothesis, which implies that estimators and tests based on such score functions are relatively robust to misspecifications of the conditional mean outcomes, see \citet{doubleML}. This robustness is particularly advantageous when estimating the conditional mean outcomes using machine learning techniques, which often entail regularization bias in estimation. 

We estimate the target parameter \( \theta \) using cross-fitting, as outlined in \citet{doubleML}. Cross-fitting estimates the models for the conditional mean outcomes and the score function \eqref{score2} using separate observations from the sample, mitigating overfitting. We assume an i.i.d. sample with \( n \) subjects, where \( i \) ranges from 1 to \( n \), representing the index of subjects in the sample. The subjects are randomly divided into \( K \) subsamples or folds of size \( n/K \), which is assumed to be an integer for simplicity. Let \( I_{k_i} \) with \( k_i \) ranging from 1 to \( K \) denote the fold containing subject \( i \), and \( I_{k_i}^c \) denote its complement, consisting of all remaining folds excluding subject \( i \).
For any subject \( i \) in fold \( k_i \), we obtain predictions \( \hat{\eta}_1^{k_i}(V_i) \) and \( \hat{\eta}_2^{k_i}(V_i) \) of the true conditional means \( \eta_1(V_i) \) and \( \eta_2(V_i) \) respectively, based on estimating the model parameters (e.g., coefficients in a lasso regression) of the conditional means in the complement set \( I_{k_i}^c \). The cross-fitted estimator for \( \theta \) is given by
\begin{align}\label{est}
	\hat{\theta}=\frac{1}{3n}\sum_{i=1}^n \begin{pmatrix}
\left(\hat{\mu}_Y^{k_i}(D_i,X_i)-\hat{\mu}_Y^{k_i}(D_i,X_i,Z_{1,i})\right)^2+\zeta_i\\
+\left(\hat{\mu}_M^{k_i}(D_i,X_i)-\hat{\mu}_M^{k_i}(D_i,X_i,Z_{1,i})\right)^2+\zeta_i\\
+\left(\hat{\mu}_Y^{k_i}(D_i,M_i,X_i)-\hat{\mu}_Y^{k_i}(D_i,M_i,X_i,Z_{2,i})\right)^2+\zeta_i
\end{pmatrix},
\end{align}
with  $(\hat{\mu}_Y^{k_i}(D_i,X_i), \hat{\mu}_M^{k_i}(D_i,X_i), \hat{\mu}_Y^{k_i}(D_i,M_i,X_i))'=\hat{\eta}_1^{k_i}(V_i)$\\ and $(\hat{\mu}_Y^{k_i}(D_i,X_i,Z_{1,i}), \hat{\mu}_M^{k_i}(D_i,X_i,Z_{1,i}), \hat{\mu}_Y^{k_i}(D_i,M_i,X_i,Z_{2,i}))'=\hat{\eta}_2^{k_i}(V_i)$.

Although \citet{huberkueck2022} primarily consider a binary instrument, their results suggest that under the null hypothesis \( H_0 \), the estimator \( \hat{\theta} \) is asymptotically normally distributed, provided certain regularity conditions hold, such as a convergence rate of the estimators \( \hat{\eta}_1^{k_i}(V_i) \) and \( \hat{\eta}_2^{k_i}(V_i) \) satisfying \( o(n^{-1/4}) \). Appendix \ref{Neymanorth} provides a formal proof for the Neyman orthogonality of score function $\phi(V,\theta,\eta)$ and the asymptotic normality of testing under the null hypothesis, along with the required regularity conditions. It is important to note that under the alternative hypothesis, the estimator is non-normal. Regarding the variance of \( \hat{\theta} \), we note that for any subject \( i \), the three squared differences adjusted by \( \zeta_i \) within the brackets of expression \eqref{est} are not independent of each other. Consequently, the conventional variance formula of the score function \eqref{score2}, \( E[(\eta_1(V)-\eta_2(V))^4]+\sigma_\zeta^2 \), is inappropriate for inference under the null hypothesis. For this reason, we apply cluster-robust variance estimation by clustering at the subject level to address this issue. 

\section{Simulations}\label{simulations}

This section evaluates the finite sample performance of our proposed testing approach through a comprehensive simulation study. We focus on assessing the robustness of the method under varying levels of confounding, instrument validity, and sample sizes, which are critical factors in applied settings where causal identification assumptions are challenging to verify. Our first simulation design replicates the causal structure described in Section \ref{Assumptions1}, specified as follows: 
\begin{eqnarray}\label{sim1}
	D &=& I\{X'\beta + 0.5 Z_1 + U_1>0\},\\
	M &=& 0.5 D + 0.5 Z_2 + X'\beta  + \delta U_1 + U_2,\notag \\
	Y &=& D + 0.5 M + X'\beta + \gamma Z_1 + \gamma Z_2 + \delta U_1  + U_3,\notag \\
	X &\sim& \mathcal{N}(0,\sigma^2_X), Z_1 \sim \mathcal{N}(0,1),  Z_2 \sim \mathcal{N}(0,1),\notag \\
	U_1&\sim& \mathcal{N}(0,1), U_2\sim \mathcal{N}(0,1),  U_3\sim \mathcal{N}(0,1), \notag
\end{eqnarray}
with $X$, $Z_1$, $Z_2$, $U_1$, $U_2$, and $U_3$ being independent of each other.
The binary treatment variable $D$ is generated as an indicator function $I(\cdot)$ based on a linear combination of pre-treatment covariates $X$, given that the coefficient vector $\beta$ is nonzero, the first instrument $Z_1$, and an unobserved error term $U_1$.
The mediator $M$ is specified as a linear function of $D$, $Z_2$, and $X$, alongside unobserved terms $U_1$ and $U_2$, allowing for potential confounding between the treatment, mediator, and outcome if $\delta \neq 0$. The outcome variable $Y$ is a linear function of treatment $D$, mediator $M$, covariates $X$ (if $\beta \neq 0)$, instruments $Z_1$ and $Z_2$ (if coefficient $\gamma\neq0$), unobservable $U_3$, and unobservable $U_1$ (if $\delta\neq 0$). The causal model implies that the (total) treatment effect is $1+0.5\cdot0.5=1.25$, the direct effect (not operating via the mediator $M$) is $1$, and the indirect (of $D$ on $Y$ via $M$) is $0.5\cdot0.5=0.25$. 
The unobserved terms $U_1, U_2, U_3$ and the instruments $Z_1, Z_2$ are standard normally distributed and independent of each other and of $X$. 
$X$ is a vector of normally distributed covariates with zero means and a covariance matrix $\sigma^2_X$, where the covariance of the $i$th and $j$th covariate in $X$ corresponds to $0.5^{|i-j|}$. The coefficient vector $\beta$ gauges the effects of the covariates on $Y$, $M$, and $D$, quantifying the degree of confounding due to observables. The $i$th element of $\beta$ is set to $0.5/i^2$ for $i=1,\ldots,p$, implying a quadratic decay in the relevance of any additional covariate $i$ for confounding.

Our study comprises 1000 simulations for each of two distinct sample sizes \( n \) consisting of 1000 and 4000 observations respectively, with the number of covariates \( X \) set to 200. To test the null hypothesis \eqref{H0_quadratic}, we estimate the conditional means involved in the score function \eqref{score2} using the \texttt{testmedident()} command in the \textsf{R} package \texttt{causalweight} by \citet{BodoryHuber2018}. We utilize default options but set $k = 2$, using 2-fold cross-fitting. We generate the random variable \( \zeta \) from a mean-zero normal distribution, \( \mathcal{N}(0,\sigma_\zeta^2) \), with the standard deviation \( \sigma_\zeta \) being inversely proportional to the sample size, \( \sigma_\zeta = 500/n \). This choice ensures that the influence of \( \zeta \) on the estimated violations becomes asymptotically negligible. Furthermore, we estimate the total as well as the natural direct and indirect treatment effects using Double Machine Learning (DML) with cross-fitting, employing the default options of the \texttt{medDML()} command in the \textsf{R} package \texttt{causalweight} by \citet{BodoryHuber2018}. This method applies lasso regression to estimate the outcome, mediator, and treatment equations, and requires that selection-on-observables holds for the treatment and the mediator (given \( X \) and given \( D,X \) respectively), as imposed in expression \eqref{joint} and Assumption \ref{A7}. The procedure drops observations with conditional treatment probabilities, known as propensity scores, close to zero or one (i.e., smaller than a threshold of 0.01 or 1\% or larger than 0.99 or 99\%) from the estimation to prevent inflation of the propensity score-based weights, which could lead to increased variance in effect estimation.

The parameters $\delta$ and $\gamma$ allow us to systematically introduce violations of the selection-on-observables assumption and instrument validity, respectively. Setting the coefficients $\delta=0$ and $\gamma=0$ in the simulations implies the satisfaction of any conditional independence assumptions imposed on the treatment, the mediator, or the instruments, such that the testable implications in Theorem \ref{mainsetup} all hold. In contrast, when \( \delta \neq 0 \), the variable \( U_1 \) becomes an unobserved confounder jointly affecting the treatment, mediator, and outcome variables. This violates the selection-on-observables assumptions on the treatment and the mediator. Furthermore, when \( \gamma \neq 0 \), the instruments exhibit direct effects on the outcome, violating instrument validity. Table \ref{tab:simulations} presents the results of our simulations under various choices of \( \delta \) and \( \gamma \). It includes the test's rejection rate (rej.\ rate) at the 5\% level of statistical significance and its average p-value (mean pval). Additionally, we report the absolute biases (bias) and root mean squared errors (RMSE) of the DML estimator of the total, direct, and indirect effects to assess how violations of the testable implications impact effect estimation performance. These statistics are provided for the direct effects under both treatment and control (dir.(1), dir.(0)), defined as \( E[Y(1,M(1))-Y(0,M(1))] \) and \( E[Y(1,M(0))-Y(0,M(0))] \), as well as for the indirect effects under both treatment and control (indir.(1), indir.(0)), defined as \( E[Y(1,M(1))-Y(1,M(0))] \) and \( E[Y(0,M(1))-Y(0,M(0))] \). However, it is worth noting that in our simulation design without treatment-mediator interaction effects, the respective effects are equivalent under treatment and control.

\begin{table}
  \centering
  \begin{adjustbox}{max width=\textwidth}
  \begin{threeparttable}
  \caption{Simulations}\label{tab:simulations}
  
  \begin{tabular}{c|c c |c c c c c| c c c c c}
  \midrule
  \midrule
  \multicolumn{3}{c}{} & \multicolumn{5}{c}{bias} & \multicolumn{5}{c}{RMSE}\tabularnewline
  sample size & rej. rate & mean pval & total & dir.(1) & dir.(0) & indir.(1) & indir.(0) & total & dir.(1) & dir.(0) & indir.(1) & indir.(0)\tabularnewline
  \midrule
  \multicolumn{13}{c}{$\delta$ = 0 \& $\gamma$ = 0}\tabularnewline
  \midrule 
  1000 & 0.039 & 0.515 & 0.005 & 0.004 & 0.004 & 0.001 & 0.001 & 0.084 & 0.079 & 0.079 & 0.052 & 0.051 \tabularnewline
  4000 & 0.048 & 0.509 & 0.001 & 0.001 & 0.000 & 0.001 & 0.002 & 0.041 & 0.037 & 0.037 & 0.022 & 0.023\tabularnewline
  \midrule 
  \multicolumn{13}{c}{$\delta=$ 1 \& $\gamma$ = 0}\tabularnewline
  \midrule
  1000 & 0.399 & 0.220 & 1.318 & 0.928 & 0.929 & 0.389 & 0.390 & 1.324 & 0.937 & 0.937 & 0.403 & 0.404\tabularnewline
  4000 & 1.000 & 0.000 & 1.376 & 0.916 & 0.914 & 0.461 & 0.459 & 1.377 & 0.918 & 0.916 & 0.464 & 0.462 \tabularnewline
  \midrule
  \multicolumn{13}{c}{$\delta$ = 0 \& $\gamma$ = 0.2}\tabularnewline
  \midrule
  1000 & 0.089 & 0.443 & 0.143 & 0.104 & 0.104 & 0.039 & 0.039 & 0.168 & 0.132 & 0.132 & 0.069 & 0.069 \tabularnewline
  4000 & 1. & 0. & 0. & 0. & 0. & 0. & 0. & 0. & 0. & 0. & 0. & 0.\tabularnewline
  \midrule
  \midrule
  \end{tabular}
  \begin{tablenotes}
    \small
    \item Notes: Column `rej. rate' gives the empirical rejection rate when setting the level of statistical significance to 0.05. `mean pval' gives the average p-value across simulations. Columns `total', `dir.(1)', `dir.(0)', `indir.(1)', `indir.(0)' provide the average absolute bias or RMSE across all samples for the total effect, the direct effects under treatment (1) and control (0), as well as the indirect effects under treatment (1) and control (0), respectively. 
  \end{tablenotes}
  \end{threeparttable}
  \end{adjustbox}
\end{table}

The top panel of Table \ref{tab:simulations} shows the results for \( \delta = 0 \) and \( \gamma = 0 \), ensuring satisfaction of all testable implications. The test maintains close adherence to the nominal 5\% significance level, with rejection rates of 4.4\% and 4.7\% for sample sizes of \(n = 1000 \) and \(n = 4000\), respectively. These results indicate that the test performs well in finite samples, minimizing Type I errors when the identifying assumptions hold. Furthermore, the decreasing bias and RMSE as sample size increases underscore the consistency of the proposed method. Correspondingly, the average p-value across the 1000 simulations is relatively high, slightly exceeding 50\%. Regarding effect estimation, we find that the absolute biases of the total, direct, and indirect effects are generally close to zero and decrease with increasing sample size. Additionally, the RMSE decreases as the sample size increases from \( n = 1000 \) to \( n = 4000 \), decaying roughly at a rate proportional to \( \sqrt{n} \). 
The intermediate panel reports the results for \( \delta = 1 \) and \( \gamma = 0 \), indicating a violation of the selection-on-observables assumptions. Accordingly, the rejection rate of our test increases to 68.8\% under the smaller sample size and reaches 100\% under \( n = 4000 \), while the average p-value decreases from 12.2\% to 0\%, which points to a very decent power of our test in the scenario considered. As expected, the DML estimator of the total, direct, and indirect effects performs poorly under both sample sizes, with absolute biases and RMSEs substantially different from zero. 

When \( \delta = 0 \) and \( \gamma = 0.2 \), violating instrument validity through moderate direct effects on the outcome, the test's power increases markedly from 8.6\% with \( n = 1000 \) to 100\% with \( n = 4000 \). Regarding the estimated effects, we note that the absolute biases are generally not substantial but not negligible either, and they only diminish slightly as the sample size increases. This phenomenon occurs because the instruments confound the treatment-outcome and mediator-outcome associations due to their direct effects on the outcome, which is not accounted for in the effect estimations that only consider \( X \) as covariates to be controlled for. Overall, the simulation results suggest that the test exhibits satisfactory size and power to detect violations of the testable implications, which increase with sample size.

Our second simulation design explores a scenario where the first instrument, $Z_1$, exerts a direct influence on the second instrument, $Z_2$. This structural adjustment is reflected by redefining $Z_2$ in model \eqref{sim1} as follows, while the remaining variable definitions are the same as before: 
\begin{align*}
Z_2 = U_4 + 0.5 Z_1, \quad U_4 \sim \mathcal{N}(0,1).
\end{align*}
Table~\ref{tab:simulations2} presents the results of 1000 simulations when testing the implications of Theorem \ref{altsetup}, which now partly contain $Z_2$ in the conditioning set. In the scenario where all testable implications are met ($\delta$ = 0 \& $\gamma$ = 0), as depicted in the top panel, the empirical size of the test closely mirrors the nominal size of 5\%, while the average p-value is larger than 50\%. Notably, the absolute biases of effect estimates remain close to zero across both sample sizes of $n = 1,000$ and $n = 4,000$ and the RMSE follows a decay rate roughly proportional to $\sqrt{n}$. Transitioning to the intermediate panel with $\delta = 1$ and $\gamma = 0$, implying a violation of the selection-on-observables assumptions, testing power quickly increases in the sample size, while the effect estimators are substantially biased. In the bottom panel with $\delta = 0$ and $\gamma = 0.2$, implying that the instruments directly affect the outcome, the test promptly gains power in the sample size, while the effect estimators are moderately biased. These findings are qualitatively similar to those of the initial simulation design discussed above.

\begin{table}

  \centering
  \begin{adjustbox}{max width=\textwidth}
  \begin{threeparttable}
  \caption{Simulations: Alternative Identifying Assumptions}\label{tab:simulations2}
  
  \begin{tabular}{c|c c |c c c c c| c c c c c}
  \midrule
  \midrule
  \multicolumn{3}{c}{} & \multicolumn{5}{c}{Bias} & \multicolumn{5}{c}{RMSE}\tabularnewline
  sample size & rej. rate & mean pval & total & dir.(1) & dir.(0) & indir.(1) & indir.(0) & total & dir.(1) & dir.(0) & indir.(1) & indir.(0)\tabularnewline
  \midrule
  \multicolumn{13}{c}{$\delta$ = 0 \& $\gamma$ = 0}\tabularnewline
  \midrule 
  1000 & 0.042 & 0.514 & 0.012 & 0.005 & 0.005 & 0.007 & 0.007 & 0.085 & 0.081 & 0.081 & 0.051 & 0.052\tabularnewline
  4000 & 0.049 & 0.510 & 0.004 & 0.001 & 0.000 & 0.005 & 0.006 & 0.040 & 0.039 & 0.039 & 0.022 & 0.023\tabularnewline
  \midrule 
  \multicolumn{13}{c}{$\delta=$ 1\& $\gamma$ = 0}\tabularnewline
  \midrule
  1000 & 0.297 & 0.286 & 1.244 & 0.914 & 0.917 & 0.327 & 0.330 & 1.250 & 0.925 & 0.927 & 0.344 & 0.347\tabularnewline
  4000 & 1.000 & 0.000 & 1.268 & 0.864 & 0.864 & 0.404 & 0.403 & 1.269 & 0.867 & 0.866 & 0.407 & 0.407\tabularnewline
  \midrule
  \multicolumn{13}{c}{$\delta$ = 0 \& $\gamma$ = 0.2}\tabularnewline
  \midrule
  1000 & 0.234 & 0.318 & 0.112 & 0.124 & 0.123 & 0.010 & 0.011 & 0.140 & 0.149 & 0.148 & 0.056 & 0.057\tabularnewline
  4000 & 1.000 & 0.000 & 0.117 & 0.123 & 0.122 & 0.005 & 0.006 & 0.124 & 0.129 & 0.128 & 0.022 & 0.023\tabularnewline
  \midrule
  \midrule
  \end{tabular}
  \begin{tablenotes}
    \small
    \item Notes: Column `rej. rate' gives the empirical rejection rate when setting the level of statistical significance to 0.05. `mean pval' gives the average p-value across simulations. Columns `total', `dir.(1)', `dir.(0)', `indir.(1)', `indir.(0)' provide the average absolute bias or RMSE across all samples for the total effect, the direct effects under treatment (1) and control (0), as well as the indirect effects under treatment (1) and control (0), respectively. 
  \end{tablenotes}
  \end{threeparttable}
  \end{adjustbox}
\end{table}

\section{Empirical application}\label{empirical}

This section presents an empirical application of our test using administrative data from Slovakia, previously analyzed by \citet{almp} to evaluate a Slovak active labor market intervention known as the `Youth Guarantee'. The dataset contains comprehensive records from the Slovak public employment service centers, which are part of the Central Office for Labour, Social Affairs, and Family of the Slovak Republic (COLSAF). These records include detailed information on unemployment history, socio-economic characteristics, participation in labor market programs, and regional characteristics.

\begin{figure}[!htp]
  \centering \caption{\label{timeline}  Timeline of interventions and variable measurements \bigskip}
  \centering 
  \includegraphics[width=0.9\textwidth]{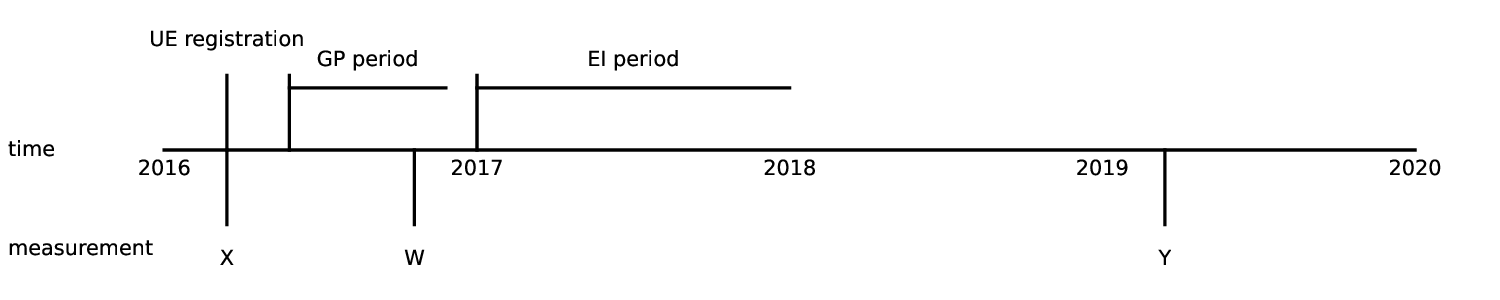}
\end{figure}

We focus on the dynamic effects of a specific sequential labor market intervention aimed at unemployed youth who registered as unemployed at the beginning of 2016. The program sequence starts with a three to six-month training called graduate practice (GP), designed to foster early-career workplace experience, albeit with less substantial monetary contributions than the subsequent interventions. The time lag between unemployment registration and the start of GP must be at least one month by law. After completing GP, participants could advance to a second program that is typically twelve months long and provides employment incentives (EI) that combine hiring incentives with subsidized employment, by offsetting up to 75\% of the costs for employing unemployed youth over twelve months, which is then followed by a compulsory employment term of six months. \citet{svabova_kramarova_2021} and \citet{stefanik_can_2020} find these interventions to have moderate but statistically significant positive effects on employment, but negative effects on earnings. In our analysis, the first intervention, $D$, is defined as receiving GP within the first 6 months after the unemployment registration, while the second intervention, $M$, is defined as receiving employment incentives within months seven to twelve since unemployment registration. The timeline of the interventions is depicted in Figure~\ref{timeline}.

Central to our testing approach are two instrumental variables, $Z_1$ and $Z_2$, designed to capture the local availability of programs $D$ and $M$, respectively, at public employment service (PES) centers. The first instrument $Z_1$ is computed based on the ratio of jobseekers enrolled in intervention $D$ in the previous year (2015) to the total influx of new jobseekers at the respective PES office during the previous two years (2014 and 2015). An analogous methodology is applied to compute the second instrument $Z_2$ related to intervention $M$. Our approach follows a well-established tradition in the literature, where local program availability is used as an instrument for participation in labor market interventions. Examples include \citet{froelich_lechner_2010}, \citet{lechner_2013}, \citet{boockmann_2014}, \citet{markussen_2014}, \citet{FrolichLechner2015}, and, in particular, \citet{caliendo_2017}, \citet{dauth_2020}, and \citet{lang_2022}, who use a similar approach as ours by comparing the number of participants in a program to the overall number of eligible jobseekers. By applying this approach in the Slovak context, our instruments aim to quantify the local propensity for receiving a specific labor market intervention. It is worth noting that the application to a particular intervention needs to be submitted by the jobseeker themselves, while it is the regional PES management that makes the final decision of acceptance, as a function of budgetary restrictions and caseworker recommendations. As such, program enrollment variability is significantly shaped by regional factors, such as unemployment rates, proximity to PES offices, and demographic considerations, including the proportion of the Roma population in the area, which we can observe and control for.

In addition to regional control variables, our data set contains rich individual-level information. The pre-treatment covariates  ($X$) are measured prior to the GP intervention (\(D\)) and consist of 264 variables that include, besides regional information, a jobseeker's marital status, the presence of dependents, comprehensive measures of education and skills, employment histories, prior claims of unemployment benefits,  willingness to relocate for work, health information, and caseworker assessments of employability prospects. We also include five post-treatment covariates ($W$) in our analysis, which are measured after (and may be influenced by) \(D\) but prior to the start of the EI intervention (\(M\)) and might potentially affect both EI participation and the outcome variable. The variables \(W\) include information on a jobseeker's participation in other labor market programs than GP ($D$) during the first intervention period and whether the jobseeker was absent from the unemployment register in months four to six (e.g., due to being employed). Finally, our outcome variable (\(Y\)) is a binary indicator for employment three years after the initial unemployment registration. The causal framework considered for testing relies on the assumptions and testable implications discussed in Section \ref{Assumptions3}, related to the illustration in Figure \ref{dag_5}. 

Our evaluation sample comprises 12,436 individuals aged 15 to 29 who registered as unemployed in 2016, remained jobless for a minimum of three months, and were tracked for up to 48 months after registration. Among them, 2,491 participated solely in GP ($D$), 1,597 solely in EI ($M$), 2,025 in both programs, and 6,323 in neither. In line with suggestions by \citet{doubleML}, Table~\ref{tab:application} displays the median p-value (pval) from 21 runs of our testing procedure outlined in Section~\ref{testing}, using the \texttt{testmedident()} command in the \textsf{R} package \texttt{causalweight} by \citet{BodoryHuber2018}. We use default options, except for the number of cross-folds, which we set to 10 instead of the default value 3. This procedure tests the implications of Theorem~\ref{dynsetup} detailed in Section~\ref{Assumptions3}. The median p-value for these runs is 24.19\%, the related test statistics 0.000421, and the standard error 0.000360, such that we cannot reject the null hypothesis of the joint satisfaction of the testable implications at conventional significance levels. Furthermore, we note that the instruments $Z_1$ and $Z_2$ exhibit highly statistically significant first-stage associations with $D$ and $M$, respectively. When utilizing the {\tt DoubleMLData()} command (with default options) in the {\tt DoubleML}  package by \cite{Bachetal2024} to estimate the first-stage associations of $D$ and $Z_1$ conditional on $X$ and of $M$ and $Z_2$ conditional on $(D,X,W)$ based on DML with lasso regression models for the instruments and interventions, the p-values are below 1\%. This suggests the satisfaction of Assumptions \ref{A4} and \ref{A5m}, which are important for the power of the test.

\begin{table}
	\centering
	\begin{adjustbox}{max width=\textwidth}
		\begin{threeparttable}
			\caption{Empirical Application}\label{tab:application}
			
			\begin{tabular}{c c c c c c c}
				\midrule
				\midrule
				teststat & se &  pval & effect & effect\_se & effect\_pval & effect\_ntrimmed \tabularnewline
				\midrule
				0.00042 & 0.00036 & 0.24189 & 0.0855 & 0.0249 & 0.0006 & 6,288 \tabularnewline  
				\midrule
				\midrule
			\end{tabular}
			\begin{tablenotes}
				\small
				\item Notes: Columns `teststat', `se', and `pval' provide the median value of the p-value and their corresponding test statistic and standard error, across 21 runs using 21 different seeds. `effect', `effect\_se', `effect\_pval' provide the dynamic treatment effect estimate, its standard error, and its p-value, respectively. `effect\_ntrimmed' gives the number of discarded observations, due to extreme propensity scores.
			\end{tablenotes}
		\end{threeparttable}
	\end{adjustbox}
	
\end{table}

Due to the insignificant test result, we suspect that the selection-on-observables assumptions regarding $D$ and $M$ are (close to being) satisfied. For this reason, Table~\ref{tab:application} also presents an estimate of the ATE (effect) of participating in both GP and EI versus not participating in any program, $E[Y(1,1)-Y(0,0)]$, along with the standard error of the effect estimate (effect\_se) and its p-value (effect\_pval). Estimation is based on the {\tt dyntreatDML} command (with default options) of the {\tt causalweight} package, which employs cross-fitted DML and lasso regression for the estimation of the models for $D$, $M$, and $Y$, as discussed in \cite{BodoryHuberLaffers}. The ATE estimate suggests that participating in both programs increases the employment probability by 8.55 percentage points, and this effect is highly statistically significant, with a p-value of less than 0.1\%. 

Table~\ref{tab:application} also reports the number of discarded observations due to extreme propensity scores related to $D$ and/or $M$ (effect\_ntrimmed), indicating that the product of the propensities to receive $D$ and $M$ is either smaller than 1\% or larger than 99\%. Specifically, 6,280 observations, or 50.5\% of the original sample, were discarded, implying that only roughly half of the individuals in the sample are sufficiently similar in terms of their covariates across different interventions to be considered for effect estimation. It is noteworthy that various propensity score thresholds for discarding observations, such as smaller than 0.5\% and larger than 99.5\%, or smaller than 5\% and larger than 95\%, yield very similar effect estimates.



A further analysis was conducted to examine the robustness of our test results to changes in the set of control variables. Specifically, the number of covariates included in the model was reduced by excluding post-treatment variables $W$ and substantially reducing $X$ to only include age and four binary indicators representing the highest level of educational attainment. This revised specification yielded a median test statistic of 0.00066, with a corresponding standard error of 0.00036 and a p-value of 6.92\%. Therefore, our test rejects the null hypothesis at the 10\% significance level, which is in line with our impression that the assumption of sequential ignorability does not appear plausible under such a limited set of control variables.

\section{Conclusion}\label{conclusion}

This paper introduces a new method for testing the identification of causal effects in mediation, dynamic treatment, and sample selection models, by making use of covariates to be controlled for and suspected instruments. It establishes sufficient testable conditions for identifying such effects in observational data. These conditions jointly imply the exogeneity of the treatment and the mediator conditional on observed variables, as well as the validity of the instruments for the treatment and mediator (or sample selection indicator). 
We explore various causal frameworks and testable implications, contingent upon whether the covariates to be controlled for are solely pre-treatment or also include post-treatment variables, and depending on the causal association between the instruments for the treatment and the mediator. We suggest a machine learning-based testing approach that handles high-dimensional covariates in a data-adaptive manner and demonstrates a decent performance in our simulation study. Furthermore, we apply the test to evaluate a sequence of active labor market programs in Slovakia, incorporating both pre- and post-treatment control variables and do not reject the testable implications for the identification of dynamic treatments when employing the local availability of labor market programs as instruments.

\newpage
\setlength\baselineskip{14.0pt}
\bibliographystyle{econometrica}
\bibliography{research}

\newpage
{\large \renewcommand{\theequation}{A-\arabic{equation}} %
\setcounter{equation}{0} \appendix
}
\appendix \numberwithin{equation}{section}
{\small
\section{Appendix}

\subsection{Proofs of identification results}\label{proofsident}

We provide a compact overview of all the identifying assumptions, testable restrictions, and results that are in the paper.

\noindent{\bf Assumptions:}
{\allowdisplaybreaks
\begin{align}
 \tag{1}M(y) &=M, D(m, z_2, y)=D, X(d, m, z_2, y)=X,  Z_1(d, m, z_2, y)=Z_1, 
  \notag \\
 Z_2(m, y) &=Z_2, \ \  \notag \forall d \in \mathcal{D}, m \in \mathcal{M},  z_2 \in \mathcal{Z}_2  \textrm{ and } y  \in \mathcal{Y}, \\ \notag 
  &\textrm{ (Causal faithfulness) + (SEM representation). }
 \\
  \tag{1m}M(y) &=M, D(m, z_2, y)=D, X(d, m, z_2, y)=X,  Z_1(d, m, z_2, y)=Z_1, 
  \notag \\
 Z_2(m, y) &=Z_2,  W(m) =W, \ \  \notag \forall d \in \mathcal{D}, m \in \mathcal{M},  z_2 \in \mathcal{Z}_2  \textrm{ and } y  \in \mathcal{Y}, \\ \notag 
  &\textrm{ (Causal faithfulness) + (SEM representation). }
 \\
 \tag{4}D & \not\!\perp\!\!\!\perp Z_1|X=x \quad \forall  x \in \mathcal{X}
 \\
  \tag{5}M & \not\!\perp\!\!\!\perp Z_2|D=d, X=x  \quad \forall d \in \mathcal{D}, x \in \mathcal{X} 
  \\
    \tag{5m}M & \not\!\perp\!\!\!\perp Z_2|D=d, X=x, W=w  \quad \forall d \in \mathcal{D}, x \in \mathcal{X}, w \in \mathcal{W}  
  \\
  \tag{6a}Y(d',m) & {\perp\!\!\!\perp} D | X=x \quad \forall d' \in \mathcal{D}, m \in \mathcal{M} \textrm{ and } x \in \mathcal{X} 
  \\
  \tag{6b}M(d) & {\perp\!\!\!\perp} D | X=x \quad \forall  d \in \mathcal{D} \textrm{ and } x \in \mathcal{X} 
  \\
  \tag{6am} Y(d',m) & {\perp\!\!\!\perp} D |  {\color{black}Z_2 = z_2}, X=x \quad \forall d', d \in \mathcal{D}, m \in \mathcal{M},  z_2 \in \mathcal{Z}_2 \textrm{ and }  x \in \mathcal{X}
  \\
  \tag{6bm}M(d) & {\perp\!\!\!\perp} D | Z_2 = z_2, X=x  \quad \forall d \in \mathcal{D}, z_2 \in \mathcal{Z}_2  \textrm{ and }  x \in \mathcal{X}
  \\
  \tag{7}Y(d',m) & {\perp\!\!\!\perp} M | D=d, X=x \quad \forall d', d \in \mathcal{D}, m \in \mathcal{M} \textrm{ and } x \in \mathcal{X}  
  \\ 
  \tag{7m}Y(d',m) & {\perp\!\!\!\perp} M | D=d, X=x, {\color{black}W=w} \quad \forall d', d \in \mathcal{D}, m \in \mathcal{M},  x \in \mathcal{X}  \textrm{ and } w \in \mathcal{W}
  \\ 
  \tag{8a}Y(d',m) & {\perp\!\!\!\perp} Z_1 | X=x \quad  \forall d' \in \mathcal{D}, m \in \mathcal{M} \textrm{ and } x \in \mathcal{X} 
  \\
  \tag{8b}M(d) & {\perp\!\!\!\perp} Z_1 | X=x \quad  \forall d \in \mathcal{D} \textrm{ and } x \in \mathcal{X} 
  \\
  \tag{8am}Y(d',m) & {\perp\!\!\!\perp} Z_1 |  {\color{black}Z_2 = z_2}, X=x \quad  \forall d' \in \mathcal{D}, m \in \mathcal{M}, z_2 \in \mathcal{Z}_2 \textrm{ and } x \in \mathcal{X} 
  \\
  \tag{8bm}M(d) & {\perp\!\!\!\perp} Z_1 | {\color{black}Z_2 = z_2}, X=x \quad  \forall d \in \mathcal{D}, z_2 \in \mathcal{Z}_2 \textrm{ and }  x \in \mathcal{X}
  \\
   \tag{9}Y(d',m) & {\perp\!\!\!\perp} Z_2 | D=d,  X=x \quad \forall d', d \in \mathcal{D}, m \in \mathcal{M} \textrm{ and } x \in \mathcal{X} 
   \\
   \tag{9m}Y(d',m) & {\perp\!\!\!\perp} Z_2 | D=d,  X=x, {\color{black}W=w} \quad \forall d', d \in \mathcal{D}, m \in \mathcal{M},  x \in \mathcal{X}  \textrm{ and } w \in \mathcal{W}
\end{align}
}

{\bf Testable implications:}
\begin{align}
   \tag{TIa} Y & {\perp\!\!\!\perp} Z_1 | D=d, X=x \quad \forall d \in \mathcal{D} \textrm{ and } x \in \mathcal{X}.   \\
   \tag{TIam}Y & {\perp\!\!\!\perp} Z_1 | D=d, {\color{black}Z_2=z_2}, X=x  \quad \forall d \in \mathcal{D}, z_2 \in \mathcal{Z}_2 \textrm{ and } x \in \mathcal{X} .
   \\
	 \tag{TIb}M & {\perp\!\!\!\perp} Z_1 | D=d, X=x\quad \forall d \in \mathcal{D} \textrm{ and } x \in \mathcal{X}.
  \\
  \tag{TIbm}M & {\perp\!\!\!\perp} Z_1 | D=d, {\color{black}Z_2=z_2 }, X=x  \quad \forall d \in \mathcal{D}, z_2 \in \mathcal{Z}_2 \textrm{ and } x \in \mathcal{X} .
  \\
   \tag{TIc}Y & {\perp\!\!\!\perp} Z_2 | D=d, M=m, X=x \quad  \forall d \in \mathcal{D}, m \in \mathcal{M} \textrm{ and } x \in \mathcal{X} 
   \\
   \tag{TId}Y & {\perp\!\!\!\perp} Z_1 | D=d, M=m, X=x \quad  \forall d \in \mathcal{D}, m \in \mathcal{M} \textrm{ and } x \in \mathcal{X} 
   \\
    \tag{TIe}Y & {\perp\!\!\!\perp} Z_2 | D=d, M=m, X=x, {\color{black} W=w}\quad \forall d \in \mathcal{D}, m \in \mathcal{M},  x \in \mathcal{X}  \textrm{ and } w \in \mathcal{W}
\end{align}

{\bf Propositions:}

\bigskip

{\bf Theorem \ref{mainsetup} (Main setup)}
  $$\text{Under }  (\ref{A1}), (\ref{A4}), (\ref{A5}): \  (\ref{A6}),(\ref{A7}),(\ref{A8}),(\ref{A9}) \iff (\ref{TIa}),(\ref{TIb}),(\ref{TIc})\notag $$

\bigskip
\bigskip

{\bf Lemma 1 ({\color{black}Alternative setup})}
 $$ \text{Under }  (\ref{A1}): \  (\ref{TIam}),(\ref{TIbm}),(\ref{TIc}) \implies  (\ref{TId}), \notag  \notag $$
        $$\text{Under }  (\ref{A1}): \  (\ref{TIbm}),(\ref{TIc}),(\ref{TId})  \implies (\ref{TIam}), \notag  \notag $$

{\bf Theorem \ref{altsetup} ({\color{black}Alternative setup})}
  $$  \text{Under }  (\ref{A1}), (\ref{A4}), (\ref{A5}): \  (\ref{A6m}),(\ref{A7}),(\ref{A8m}),(\ref{A9}) \iff(\ref{TIam}),(\ref{TIbm}),(\ref{TIc}) \iff (\ref{TIbm}),(\ref{TIc}),(\ref{TId}) \notag$$

\bigskip

{\bf  Theorem \ref{dynsetup} ({\color{black}Dynamic setup})}
 $$ \text{Under }  (\ref{A1m}), (\ref{A4}), (\ref{A5m}): \  (\ref{A6}),(\ref{A7m}),(\ref{A8}),(\ref{A9m}) \iff (\ref{TIa}),(\ref{TIb}),(\ref{TIe}) \notag $$

\noindent{\bf Proofs (computational approach):}

\bigskip

In what follows, $A \rightarrow B$ denotes a directed edge from $A$ to $B$ in the respective graph ($A \leftarrow B$ for other direction), and $A - B$ denotes the existence of some path between $A$ and $B$, with no restriction on edge orientations and $A \rightsquigarrow B$ denotes a directed path from $A$ to $B.$

The proofs of all these results can be done analytically following the steps similar to \citet{huberkueck2022}.

These theorems as well as all the steps in the proof can also be verified computationally. While these are not strictly necessary, we include them as they offer transparency and intuition.

We may consider all the possible DAGs with observed variables $Y, D, X, Z_1, Z_2$ (and $W$ for the dynamic setup) and with unobserved confounders for possibly all the pairs of the observed variables. We do not need to consider 
\begin{itemize}
\item[(i)] unobserved colliders, as these paths are closed anyway, 
\item[(ii)] unobserved mediators, as these can be interpreted as direct paths, 
\item[(iii)] confounders for more than two observed variables, as these are equivalent to the existence of multiple pair-wise confounders from the point of view of existence of open paths and hence identification.
\end{itemize}

Restricting to pairwise latent confounders is without loss of generality for d-separation among observed variables. This follows from the latent projection result of \citet{richardson2002ancestral} (Theorem 4.18): marginalizing a DAG over latent variables produces an acyclic directed mixed graph (ADMG) that preserves all conditional independence relations among observed variables. 

We will now translate all the assumptions into the DAG semantics.

Let $G$ denote the original graph and let $G_D$ and $G_{DM}$ denote the interventional graph in which all the arrows starting from $D$ or $\{D,M\}$ are removed. This follows from the Single World Intervention Graphs (SWIGs) framework of \citet{richardson2013single}, which provides complete rules for translation of potential outcome notation to structural causal model notation.

\bigskip

\noindent{\bf Assumptions (DAG semantics):}
{\allowdisplaybreaks
\begin{align}
 \tag{1}
&\textrm{There are no directed paths in the following directions:} \\
&  Y \rightsquigarrow M,   Y \rightsquigarrow X,    Y \rightsquigarrow Z_1,   Y \rightsquigarrow Z_2,  
 M \rightsquigarrow D,   M \rightsquigarrow X,   M \rightsquigarrow Z_1,    M \rightsquigarrow Z_2, \notag \\
& Z_2 \rightsquigarrow D,   Z_2 \rightsquigarrow X,   Z_2 \rightsquigarrow Z_1,    Z_2 \rightsquigarrow D, 
 D \rightsquigarrow X,   D \rightsquigarrow Z_1  \notag
 \textrm{ in graph } G
 \\
  \tag{1m}
&\textrm{There are no directed paths in the following directions:} \\
&  Y \rightsquigarrow M,   Y \rightsquigarrow X,    Y \rightsquigarrow Z_1,   Y \rightsquigarrow Z_2,  
 M \rightsquigarrow D,   M \rightsquigarrow X,   M \rightsquigarrow Z_1,    M \rightsquigarrow Z_2, \notag \\
& Z_2 \rightsquigarrow D,   Z_2 \rightsquigarrow X,   Z_2 \rightsquigarrow Z_1,    Z_2 \rightsquigarrow D, 
 D \rightsquigarrow X,   D \rightsquigarrow Z_1, M \rightsquigarrow W  \notag
 \textrm{ in graph } G \\
 \tag{4} D \textrm{ and } Z_1& \textrm{ are d-connected with conditioning set } \{ X \}  \textrm{ in graph } G
 \\
  \tag{5} M \textrm{ and } Z_2& \textrm{ are d-connected with conditioning set } \{ D, X \} \textrm{ in graph } G
  \\
    \tag{5m} M \textrm{ and } Z_2& \textrm{ are d-connected with conditioning set } \{ D, X, W \} \textrm{ in graph } G
\\
  \tag{6a} Y \textrm{ and } D& \textrm{ are d-separated with conditioning set } \{ X \} \textrm{ in graph } G_{DM} 
  \\
  \tag{6b} M \textrm{ and } D& \textrm{ are d-separated with conditioning set } \{ X \} \textrm{ in graph } G_{D}
\\
  \tag{6am} Y \textrm{ and } D& \textrm{ are d-separated with conditioning set } \{ X,{\color{black} Z_2} \} \textrm{ in graph } G_{DM} 
  \\
  \tag{6bm} M \textrm{ and } D& \textrm{ are d-separated with conditioning set } \{ X,{\color{black} Z_2} \} \textrm{ in graph } G_{D} \\
  \tag{7} Y \textrm{ and } M& \textrm{ are d-separated with conditioning set } \{ D,X \} \textrm{ in graph } G_{DM} 
  \\ 
  \tag{7m} Y \textrm{ and } M& \textrm{ are d-separated with conditioning set } \{ D,X,{\color{black} W} \} \textrm{ in graph } G_{DM} 
  \\ 
  \tag{8a} Y \textrm{ and } Z_1& \textrm{ are d-separated with conditioning set } \{ X \} \textrm{ in graph } G_{DM} 
  \\
  \tag{8b} M \textrm{ and } Z_1& \textrm{ are d-separated with conditioning set } \{ X \} \textrm{ in graph } G_{D} 
  \\
  \tag{8am} Y \textrm{ and } Z_1& \textrm{ are d-separated with conditioning set } \{ X,{\color{black} Z_2} \} \textrm{ in graph } G_{DM} 
  \\
  \tag{8bm} M \textrm{ and } Z_1& \textrm{ are d-separated with conditioning set } \{ X,{\color{black} Z_2} \} \textrm{ in graph } G_{D} \\
   \tag{9} Y \textrm{ and } Z_2& \textrm{ are d-separated with conditioning set } \{ D,X \} \textrm{ in graph } G_{DM} 
   \\
   \tag{9m} Y \textrm{ and } Z_2& \textrm{ are d-separated with conditioning set } \{ D,X,{\color{black} W} \} \textrm{ in graph } G_{DM}
\end{align}
{\bf Testable implications :}
\begin{align} 
   \tag{TIa}  Y \textrm{ and } Z_1 & \textrm{ are d-separated with conditioning set } \{ X,D \} \textrm{ in graph } G
   \\
   \tag{TIam} Y \textrm{ and } Z_1 & \textrm{ are d-separated with conditioning set } \{ X,D,{\color{black}Z_2} \} \textrm{ in graph } G
   \\
	 \tag{TIb} M \textrm{ and } Z_1 & \textrm{ are d-separated with conditioning set } \{ X,D \} \textrm{ in graph } G
  \\
  \tag{TIbm} M \textrm{ and } Z_1 & \textrm{ are d-separated with conditioning set } \{ X,D,{\color{black}Z_2} \} \textrm{ in graph } G
  \\
   \tag{TIc} Y \textrm{ and } Z_2 & \textrm{ are d-separated with conditioning set } \{ X,D,M \} \textrm{ in graph } G
   \\
   \tag{TId} Y \textrm{ and } Z_1 & \textrm{ are d-separated with conditioning set } \{ X,D,M \} \textrm{ in graph } G
   \\
    \tag{TIe} Y \textrm{ and } Z_2 & \textrm{ are d-separated with conditioning set } \{ X,D,M,{\color{black}W} \} \textrm{ in graph } G
\end{align}
}

We conduct an exhaustive search across all the possible DAGs. We restrict ourselves only to those for which Assumption 1 holds, which immediately removes the possibility of many of the arrows. For instance, there cannot be any direct arrow between $Y \rightarrow M.$ Then, given that all the assumptions are conditional on $X$, we do not have to consider $X$ explicitly in the DAGs but keep it implicit. This reduces the dimension of the graphs significantly and makes them computationally tractable. There are 10 different arrows between the observed variables to consider that are not ruled out by the assumption (\ref{A1}). We also consider confounders for all the pairs of observed variables, hence $\binom{5}{2} = 10$ different unobserved confounders. Altogether we have $2^{10} 2^{10} = 1048576$ DAGs to search through. We utilized the \textsf{R} packages {\tt daggity} by \cite{dagitty} and {\tt pcalg} by \cite{pcalg} to check which of the assumptions (\ref{A1}),(\ref{A1m}),(\ref{A4}),(\ref{A5}),(\ref{A6}),(\ref{A6m}),(\ref{A7}),(\ref{A7m}),(\ref{A8}),(\ref{A8m}),(\ref{A9}),(\ref{A9m})  and testable implications (\ref{TIa}),(\ref{TIam}),(\ref{TIb}),(\ref{TIbm}),(\ref{TIc}),(\ref{TId}),(\ref{TIe}) are valid. 

We used {\tt pcalg2daggity()} to translate an adjacency matrix into a DAG and {\tt dconnected()}/{\tt dseparated()} for the existence/non-existence of a path in a certain graph. Interventional graphs $G_D$ and $G_{DM}$ were obtained by removing all the arrows pointing from $D$ and $\{D,M\}$, respectively.

This computational approach provides us with a direct and principled way of proving theorems related to the testable implications regardless of the type of assumptions imposed on the causal structure. As an example, consider Theorem \ref{mainsetup}. There are 735232 DAGs that satisfy assumptions (\ref{A1}), (\ref{A4}), (\ref{A5}). Out of these
\begin{itemize}
\item[(i)] 480 DAGs satisfy (\ref{A6}), (\ref{A7}), (\ref{A8}), (\ref{A9}) and at the same time, satisfy (\ref{TIa}), (\ref{TIb}), (\ref{TIc}),
\item[(ii)] 73043 (=73523-480) DAGs that do not satisfy at least one of these assumptions: (\ref{A6}), (\ref{A7}), (\ref{A8}), (\ref{A9}) and at the same time, do not satisfy at least one of these testable implications: (\ref{TIa}), (\ref{TIb}), (\ref{TIc}).
\end{itemize}
By the mechanical exhaustive search  we establish the equivalence directly as we have complete information on which specific DAGs satisfy which assumptions/testable implications.\footnote{All the computations take about 10 hours on a standard desktop computer (mid 2023, M2 Pro processor, 32GB RAM).}

We have verified Theorems \ref{mainsetup} and \ref{altsetup} computationally. While the space of all the DAGs in Theorem \ref{dynsetup} is too large for an exhaustive search, the proof fortunately requires only a simple modification of the analytical (rather than computational) proof of Theorem \ref{mainsetup}, as provided below.

\bigskip
\bigskip
\bigskip

\noindent{\bf Proofs (analytical approach):}

\bigskip

We subsequently present the proofs of Theorems \ref{mainsetup} and \ref{dynsetup} based on the analytical (rather than computational) approach. The proof for the Theorem \ref{altsetup} follows in analogous manner and is omitted for the sake of brevity.

{\it Proof of Theorem \ref{mainsetup}}

``$\implies$ (\ref{TIa})"

By contradiction.

Let (\ref{TIa})': there exists an open path $Y - Z_1$ in $G$ conditional on $\{D,X\}$.

By (\ref{A8}(a)) there is no open path  $Y - Z_1$ in $G_{DM}$ conditional on $\{X\}$.

We may have one of the two situations:
\begin{itemize}
    \item[(i)] the conditioning on $D$ that opens up a path in $G$ conditional on $\{X\}$. 
   
    So it has to be a collider (or a descendant of a collider).

Thus we need to have $Y - ... \rightarrow D \leftarrow ... - Z_1$ in $G$ where $Y - ... \rightarrow D$ has to be open in $G$ conditional on $\{X\}.$

But $Y - D$ is closed in $G_{DM}$ conditional on $\{X\}$ by (\ref{A6}(a)).

So it is the removing of arrows from $\{D,M \}$ that makes $Y$ and $Z_1$ no longer connected. Given that there is an arrow to $D$, so it has to be removing the arrow from $M$ that makes $Y$ and $Z_1$ no longer connected.
Therefore we need to have $Y - ... - M -... \rightarrow D$, where both $Y - ... - M$ and $M -... \rightarrow D$ are open in $G$ conditional on $\{X\}$. But according to (\ref{A6}(b)), $M - D$ is closed in $G_D$ conditional on $\{X\}.$ $\text{\Lightning}.$
    \item[(ii)] removing arrows from $\{D,M\}$ closes the $Y - Z_1$ path what would otherwise be open in $G_D$.

    Given (i), we know that $D$ cannot be a collider on the path $Y - Z_1$ in $G$, in any other case conditioning on it would block the path hence (\ref{TIa})' would not hold. It therefore has to be an arrow from $M$ that is removed that causes $Y$ and $Z_1$ to be no longer connected. So we need to have $Y - ... - M -... \rightarrow Z_1$ which is open in $G$ and there is no $D$ on this path. So $M - Z_1$ is open and also $Y - M$ is open. But (\ref{A8}(b)) says $M - Z_1$ is closed in $G_D$ conditional on $\{X\}$. $\text{\Lightning}.$
\end{itemize}

``$\implies$ (\ref{TIb})"

By contradiction.

Let (\ref{TIb})': there exists an open path $M - Z_1$ in $G$ conditional on $\{D,X\}$.

By (\ref{A8}(b)) there is no open path  $M - Z_1$ in $G_{D}$ conditional on $\{X\}$.

We may have one of the two situations:
\begin{itemize}
    \item[(i)] the conditioning on $D$ opens up a path in $G$ conditional on $\{X\}$. 

    So it has to be a collider (or a descendant of a collider).

    Thus we need to have $M - ... \rightarrow D \leftarrow ... - Z_1$ in $G$ where $M - ... \rightarrow D$ has to be open in $G$ conditional on $\{X\}.$

But $M - D$ is closed in $G_{D}$ conditional on $\{X\}$ by (\ref{A6}(b)). $\text{\Lightning}.$
     
    \item[(ii)] removing arrows from $D$ closes the $M - Z_1$ path what would otherwise be open in $G$.
    
    Given (i), we know that $D$ cannot be a collider on the path $M - Z_1$ in $G$. But given (\ref{TIb})' conditioning on $D$ would block this path. $\text{\Lightning}.$

\end{itemize}

``$\implies$ (\ref{TIc})"

By contradiction.

Let (\ref{TIc})': there exists an open path $Y - Z_2$ in $G$ conditional on $\{D,M,X\}$.

By (\ref{A9}) there is no open path  $Y - Z_2$ in $G_{DM}$ conditional on $\{D, X\}$.

We may have one of the two situations:

\begin{itemize}
    \item[(i)] the conditioning on $M$ opens up a path $Y - Z_2$ in $G$ conditional on $\{D,X\}$. 

    So it has to be a collider (or a descendant of a collider).

    Thus we need to have $Y - ... \rightarrow M \leftarrow ... - Z_2$ in $G$ where $Y - ... \rightarrow M$ has to be open in $G$ conditional on $\{D,X\}.$

But $Y - M$ is closed in $G_{DM}$ conditional on $\{D,X\}$ by (\ref{A7}). $\text{\Lightning}.$
     
    \item[(ii)] removing arrows from $\{D,M\}$ closes the $Y - Z_2$ path what would otherwise be open in $G$.
    
    Given (i), we know that $M$ cannot be a collider on the path $Y - Z_2$ in $G$. But given (\ref{TIc})' conditioning on $M$ would block this path. 
    It therefore has to be an arrow from $D$ that is removed that causes $Y$ and $Z_2$ to be no longer connected.
    So we need to have $Y - ... - D -... \rightarrow Z_2$ which is open in $G$ and there is no $M$ on this path. So $Y - D$ is open and also $D - Z_2$ is open. But (\ref{A6}(a)) says $Y - D$ is closed in $G_{DM}$ conditional on $\{X\}$. $\text{\Lightning}.$

\end{itemize}

``(\ref{A8}(a)) $\impliedby$ "

By contradiction.

Let (\ref{A8}(a))': there exists an open path $Y - Z_1$ in $G_{DM}$ conditional on $\{X\}$.

By (\ref{TIa}) there is no open path  $Y - Z_1$ in $G$ conditional on $\{D, X\}$ and hence not even in $G_{DM}$ (removing arrow cannot create a new path).

Given that (\ref{A8}(a))' and (\ref{TIa}) hold simultaneously, we need to have that conditioning on $D$ closes the otherwise open path $Y - Z_1$ in $G_{DM}.$ But there are no arrows from $D$ in $G_{DM},$ so $D$ can only be a collider on the path $Y - Z_1,$ but this would violate (\ref{TIa}). $\text{\Lightning}.$

``(\ref{A6}(a)) $\impliedby$ "

By contradiction.

Let (\ref{A6}(a))': there exists an open path $Y - D$ in $G_{DM}$ conditional on $\{X\}$.

By (\ref{TIa}) there is no open path  $Y - Z_1$ in $G$ conditional on $\{D, X\}$ 

By (\ref{A4}) there is an open path $D - Z_1$  in $G$ conditional on $\{X\}$ 

Thus we need to have that conditioning on $D$ blocks the $Y - D - Z_1$ path.

We have one of the following situations.
\begin{itemize}
    \item[(i)] $Y -... \rightarrow D \leftarrow ... - Z_1$: ruled out, as $D$ cannot be a collider.
    \item[(ii)] $Y -... \leftarrow D \rightarrow ... - Z_1$: ruled out, because $Y -... \leftarrow D$ is open in $G_{DM}$ and there are not arrows from $D$ in $G_{DM}$.
    \item[(iii)] $Y -... \leftarrow D \leftarrow ... - Z_1$:  ruled out, because $Y -... \leftarrow D$ is open in $G_{DM}$ and there are not arrows from $D$ in $G_{DM}$,
     \item[(iv)] $Y -... \rightarrow D \rightarrow ... - Z_1$.
     
     Consider the path $Y -... \rightarrow D$. Because of (\ref{A1}) there is no directed path from $Y$ to $D$ and hence there has to be a confounder on this path because any collider would block the path and violate (\ref{A6}(a))'. The only possible confounder compatible with (\ref{A1}) is an unobserved variable. Therefore we could only have a path of type $Y \leftarrow U \rightarrow D \rightarrow ... - Z_1$. But because of  $D \rightarrow ... - Z_1$  (\ref{A1}) there cannot be a directed path from $D$ to $Z_1$, thus there must exist a collider on this path  $D \rightarrow ... - Z_1$. The only possible observable collider compatible with (\ref{A4}) is $X$ but a directed $X \leftarrow Z_1$ path would violate (\ref{A1}). $\text{\Lightning}.$

\end{itemize}

``(\ref{A8}(b)) $\impliedby$ "

By contradiction.

Let (\ref{A8}(b))': there exists an open path $M - Z_1$ in $G_{D}$ conditional on $\{X\}$.

By (\ref{TIb}) there is no open path  $M - Z_1$ in $G$ conditional on $\{D, X\}$ and hence not even in $G_{D}$ (removing arrow cannot create a new path).

Given that (\ref{A8}(b))' and (\ref{TIb}) hold simultaneously, we need to have that conditioning on $D$ closes otherwise open path $M - Z_1$ in $G_{D}.$ But there are no arrows from $D$ in $G_{D},$ so $D$ can only be a collider on the path $M - Z_1,$ but this would violate (\ref{TIb}). $\text{\Lightning}.$

``(\ref{A6}(b)) $\impliedby$ "

By contradiction.

Let (\ref{A6}(b))': there exists an open path $M - D$ in $G_{D}$ conditional on $\{X\}$.

By (\ref{TIb}) there is no open path  $M - Z_1$ in $G$ conditional on $\{D, X\}$ 

By (\ref{A4}) there is an open path $D - Z_1$  in $G$ conditional on $\{X\}$ 

Thus we need to have that conditioning on $D$ blocks the $M - D - Z_1$ path.

We have one of the following situations.
\begin{itemize}
    \item[(i)] $M -... \rightarrow D \leftarrow ... - Z_1$:ruled out, as $D$ cannot be a collider.
    \item[(ii)] $M -... \leftarrow D \rightarrow ... - Z_1$: ruled out, because $Y -... \leftarrow D$ is open in $G_{D}$ and there are not arrows from $D$ in $G_{D}$.
    \item[(iii)] $M -... \leftarrow D \leftarrow ... - Z_1$:  ruled out, because $Y -... \leftarrow D$ is open in $G_{D}$ and there are not arrows from $D$ in $G_{D}$,
     \item[(iv)] $M -... \rightarrow D \rightarrow ... - Z_1$.
     
     Consider the path $M -... \rightarrow D$. Because of (\ref{A1}) there is no directed path from $M$ to $D$ and hence there has to be a confounder on this path, because any collider would block the path and violate (\ref{A6}(b))'. The only possible confounder compatible with (\ref{A1}) is an unobserved variable. Therefore we could only have a path of type $M \leftarrow U \rightarrow D \rightarrow ... - Z_1$. But because of  $D \rightarrow ... - Z_1$  (\ref{A1}) there cannot be a directed path from $D$ to $Z_1$, thus there must be a collider on this path  $D \rightarrow ... - Z_1$. The only possible observable collider compatible with (\ref{A4}) is $X$ but directed $D \rightarrow X$ path would violate (\ref{A1}).$\text{\Lightning}.$

\end{itemize}

``(\ref{A9}) $\impliedby$ "

By contradiction.

Let (\ref{A9})': there exists an open path $Y - Z_2$ in $G_{DM}$ conditional on $\{D,X\}$.

By (\ref{TIc}) there is no open path  $Y - Z_2$ in $G$ conditional on $\{D, X, M\}$ and hence not even in $G_{DM}$ (removing arrow cannot create a new path).

Given that (\ref{A9})' and (\ref{TIc}) hold simultaneously, we need to have that conditioning on $M$ closes otherwise open path $Y - Z_2$ in $G_{DM}.$ But there are no arrows from $M$ in $G_{DM},$ so $M$ can only be a collider on the path $Y - Z_2,$ but this would violate (\ref{TIc}). $\text{\Lightning}.$

``(\ref{A7}) $\impliedby$ "

By contradiction.

Let (\ref{A7})': there exists an open path $Y - M$ in $G_{DM}$ conditional on $\{D, X\}$.

By (\ref{TIc}) there is no open path  $Y - Z_2$ in $G$ conditional on $\{D, X, M\}$ 

By (\ref{A5}) there is an open path $M - Z_2$  in $G$ conditional on $\{D, X\}$ 

Thus we need to have that conditioning on $M$ blocks the $Y - M - Z_2$ path.

We have one of the following situations.
\begin{itemize}
    \item[(i)] $Y -... \rightarrow M \leftarrow ... - Z_2$: ruled out, as $M$ cannot be a collider.
    \item[(ii)] $Y -... \leftarrow M \rightarrow ... - Z_2$: ruled out, because $Y -... \leftarrow M$ is open in $G_{DM}$ and there are no arrows from $M$ in $G_{DM}$.
    \item[(iii)] $Y -... \leftarrow M \leftarrow ... - Z_2$:  ruled out, because $Y -... \leftarrow M$ is open in $G_{DM}$ and there are no arrows from $M$ in $G_{DM}$,
     \item[(iv)] $Y -... \rightarrow M \rightarrow ... - Z_2$.
     
     Consider the path $Y -... \rightarrow M$. Because of (\ref{A1}) there is no directed path from $Y$ to $M$ and hence there needs to be a confounder on this path because any collider would block the path and violate (\ref{A6}(b))'. The only possible confounder compatible with (\ref{A1}) is an unobserved variable. Therefore we could only have a path of type $Y \leftarrow U \rightarrow M \rightarrow ... - Z_2$. But because of   (\ref{A1}) there cannot be a directed path from $M$ to $Z_2$, thus we need to have a collider on this path  $M \rightarrow ... - Z_2$. The only possible observable collider compatible with (\ref{A5}) is $X$ or $D$ but directed $X \leftarrow Z_2$ or $D \leftarrow Z_2$ path would violate (\ref{A1}).$\text{\Lightning}.$

\end{itemize}

This completes the proof. $\blacksquare$

{\it Proof of  Lemma 1 (a)}

Let (\ref{TId})', then the path $Y - Z_1$ is open given $\{D,X,M\}$ in $G.$

By (\ref{TIam}), there is no open path $Y-Z_1$ given $\{D,X,Z_2\}$ in $G.$

We may have one of the two situations:

\begin{itemize}
    \item[(i)] 
Path $Y-Z_1$ is open given $\{D,X\}$ in $G$,
then (i) $M$ does not close it and (ii) $Z_2$ closes it
So $Y - Z_2 - Z_1$ has to be closed by $Z_2$, but open $Y-Z_2$ path contradicts (\ref{TIc}) $\text{\Lightning}.$
    \item[(ii)] 
Path $Y-Z_1$ is closed given $\{D,X\}$ in $G$
then (i) $M$ is a collider
and (ii) $Z_2$ is not a collider
we have to have
$Y - ...\rightarrow M \leftarrow  ... -Z_1$
where $M - Z_1$ is open a this contradicts (\ref{TIb})
$\text{\Lightning}.$

\end{itemize}

\bigskip

{\it Proof of  Lemma 1 (b)}

Let (\ref{TIam})', then the path $Y - Z_1$  is open given $\{D,X,Z_2\}$ in $G.$

By (\ref{TId}), there is no open path $Y-Z_1$ given $\{D,X,M\}$ in $G.$

We may have one of the two situations:

\begin{itemize}
    \item[(i)] 
Path $Y-Z_1$ is open given $\{D,X\}$ in $G$,
then (i) $Z_2$ does not close it
and (ii) $M$ closes it. 
So $Y - M - Z_1$ has to be closed by $M$, but the open $M-Z_1$ path contradicts (\ref{TIb}) $\text{\Lightning}.$
    \item[(ii)] 
Path $Y-Z_1$ is closed given $\{D,X\}$ in $G$,
then (i)  $Z_2$ is a collider
and (ii) $M$ is not a collider;
it must hold that
$Y - \rightarrow Z_2 \leftarrow  ... -Z_1$,
where $Y - Z_2$ is open a this contradicts (\ref{TIc})
$\text{\Lightning}.$

\end{itemize}

{\it Proof of Theorem \ref{dynsetup}}

Several results in the proof of Theorem \ref{mainsetup} are unaffected by the inclusion of post-treatment covariates $W,$ and thus remain unchanged.
This applies to the following assertions:
``$\implies$ (\ref{TIa})",
``$\implies$ (\ref{TIb})",
``(\ref{A8}(a)) $\impliedby$ ",
``(\ref{A6}(a)) $\impliedby$ ",
``(\ref{A8}(b)) $\impliedby$ ",
``(\ref{A6}(b)) $\impliedby$ ".

The implications that remain to be proven are:
``$\implies$ (\ref{TIe})",
``(\ref{A7m}) $\impliedby$ ",
``(\ref{A9m}) $\impliedby$ ".
These results can be shown by modifications of the previous analytical proof, as outlined below.

``$\implies$ (\ref{TIe})"

By contradiction.

Let (\ref{TIe})': there exists an open path $Y - Z_2$ in $G$ conditional on $\{D,M,X, W\}$.

By (\ref{A9m}) there is no open path  $Y - Z_2$ in $G_{DM}$ conditional on $\{D, X, W\}$.

We may have one of the two situations:

\begin{itemize}
    \item[(i)] The conditioning on $M$ opens up a path $Y - Z_2$ in $G$ conditional on $\{D,X, W\}$, such that it is a collider (or a descendant of a collider).

    Thus it must hold that $Y - ... \rightarrow M \leftarrow ... - Z_2$ in $G$, where $Y - ... \rightarrow M$ has to be open in $G$ conditional on $\{D,X, W\}.$

But $Y - M$ is closed in $G_{DM}$ conditional on $\{D,X,W\}$ by (\ref{A7m}). $\text{\Lightning}.$
     
    \item[(ii)] Removing arrows from $\{D,M\}$ closes the $Y - Z_2$ path what would otherwise be open in $G$.
    
    Given (i),  $M$ cannot be a collider on the path $Y - Z_2$ in $G$. Furthermore, given (\ref{TIe})', conditioning on $M$ would block any path $Y - Z_2$ operating via $M$. 
    For this reason, there has to be an arrow from $D$ that is removed such that $Y$ and $Z_2$ are no longer connected. So we need to have $Y - ... - D -... \rightarrow Z_2$ which is open in $G$ and there is no $M$ on this path. So $Y - D$ is open and also $D - Z_2$ is open. But (\ref{A6}(a)) says $Y - D$ is closed in $G_{DM}$ conditional on $\{X\}$. $\text{\Lightning}.$

\end{itemize}

``(\ref{A9m}) $\impliedby$ "

By contradiction.

Let (\ref{A9m})': there exists an open path $Y - Z_2$ in $G_{DM}$ conditional on $\{D,X,W\}$.

By (\ref{TIe}) there is no open path  $Y - Z_2$ in $G$ conditional on $\{D, X, M, X\}$ and hence not even in $G_{DM}$ (removing arrow cannot create a new path).

Given that (\ref{A9m})' and (\ref{TIe}) hold simultaneously, it must hold that conditioning on $M$ closes the otherwise open path $Y - Z_2$ in $G_{DM}.$ But there are no arrows from $M$ in $G_{DM},$ so $M$ can only be a collider on the path $Y - Z_2,$ which would violate (\ref{TIe}). $\text{\Lightning}.$

``(\ref{A7m}) $\impliedby$ "

By contradiction.

Let (\ref{A7m})': there exists an open path $Y - M$ in $G_{DM}$ conditional on $\{D, X, W\}$.

By (\ref{TIe}) there is no open path  $Y - Z_2$ in $G$ conditional on $\{D, X, M, W\}$ 

By (\ref{A5m}) there is an open path $M - Z_2$  in $G$ conditional on $\{D, X, W\}$ 

Thus, it must hold that conditioning on $M$ blocks the $Y - M - Z_2$ path.

We have one of the following situations.
\begin{itemize}
    \item[(i)] $Y -... \rightarrow M \leftarrow ... - Z_2$: ruled out, as $M$ cannot be a collider.
    \item[(ii)] $Y -... \leftarrow M \rightarrow ... - Z_2$: ruled out, because $Y -... \leftarrow M$ is open in $G_{DM}$ and there are no arrows from $M$ in $G_{DM}$.
    \item[(iii)] $Y -... \leftarrow M \leftarrow ... - Z_2$: ruled out, because $Y -... \leftarrow M$ is open in $G_{DM}$ and there are no arrows from $M$ in $G_{DM}$,
     \item[(iv)] $Y -... \rightarrow M \rightarrow ... - Z_2$.
     
     Consider the path $Y -... \rightarrow M$. Because of (\ref{A1m}), there is no directed path from $Y$ to $M$ and hence there has to be a confounder on this path because any collider would block the path and violate (\ref{A6}(b))'. The only possible confounder compatible with (\ref{A1m}) is an unobserved variable. Therefore, the only possible path would be one of the type $Y \leftarrow U \rightarrow M \rightarrow ... - Z_2$. But because of   (\ref{A1m}), any directed path from $M$ to $Z_2$ is ruled out, which necessarily implies a collider on the path  $M \rightarrow ... - Z_2$. The only possible observable colliders compatible with (\ref{A5m}) would be $X$, $D$, or $W$. However, directed paths $X \leftarrow Z_2$, $D \leftarrow Z_2$, or $M \rightarrow W$ would violate (\ref{A1m}). $\text{\Lightning}.$

\end{itemize}

This completes the proof. $\blacksquare$

\subsection{Proof of Neyman orthogonality and asymptotic normality}\label{Neymanorth}

To prove the Neyman orthogonality of score function \eqref{score2} in Section \ref{testing} as well as the asymptotic normality of the testing approach under the null hypothesis \ref{H0_quadratic}, we closely follow \citet{huberkueck2022} and apply Theorem 3.1 in \cite{doubleML}. All bounds in the proof hold uniformly over $P\in\mathcal{P}$, but we omit this qualifier for the sake of brevity. We use $C$ to denote a strictly positive constant that is independent of $n$ and $P\in\mathcal{P}$. The value of $C$ may change at each appearance.

First, we note that the score in \eqref{score2} is linear
\begin{align*}
\phi_2(V,\theta,\eta)=\phi_2^a(V,\eta)\theta+\phi_2^b(V,\eta)
\end{align*}
with $\phi_2^a(V,\eta)=-1$ and $\phi_2^b(V,\eta)=(\eta_1(V)-\eta_2(V))^2+\zeta$. With a slight abuse of notation, we henceforth denote by $\theta_0$ and $\eta_0=(\eta_{1,0},\eta_{2,0})$ the true values of expression \ref{H0_theta} and the conditional means including and excluding the instruments, respectively, defined in Section \ref{testing}, while $\theta$ and $\eta=(\eta_{1},\eta_{2})$ denote possibly distinct values of those parameters. We note that the score satisfies by construction the following moment condition for the true values $\theta_0$ and $\eta_0$,
\begin{align*}
E[\phi_2(V,\theta_0,\eta_0)]=E[(\eta_{1,0}(V)-\eta_{2,0}(V))^2-\theta_0+\zeta]=0.
\end{align*}
Furthermore, under the null hypothesis $H_0$, it holds
\begin{align*}
&\partial_r E[\phi_2(V,\theta_0,\eta_0 + r(\eta-\eta_0))]\big|_{r=0}\\
&=E\left[\partial_r\phi_2(V,\theta_0,\eta_0 + r(\eta-\eta_0))\big|_{r=0}\right]\\
&=E\left[\partial_r\left(\eta_{1,0}(V)+r(\eta_1-\eta_{1,0})(V)-(\eta_{2,0}(V)+r(\eta_2-\eta_{2,0})(V))\right)^2\big|_{r=0}\right]\\
&=E\left[2\underbrace{(\eta_{1,0}(V)-\eta_{2,0}(V))}_{=0 \text{\ a.s.}}((\eta_1-\eta_{1,0})(V)-(\eta_2-\eta_{2,0})(V))\right]=0.
\end{align*}
Therefore, the score function satisfies Neyman orthogonality. 

Furthermore, the identification condition 
$J_0:=E[\phi_2^a(W,\eta)]=-1$ holds, such that Assumption 3.1 in \cite{doubleML} is satisfied. Next, we demonstrate the satisfaction of Assumption 3.2 in \cite{doubleML} to complete the proof of asymptotic normality. We define the realization set $\mathcal{T}_n$ of the conditional means as the set of all P-square-integrable functions $\eta$ such that
\begin{align*}
\|\eta_{0}-\eta\|_{P,2q}&\le C,\\
\|\eta_{0}-\eta\|_{P,4}&\le \delta_n,\\
\|\eta_{0}-\eta\|_{P,2}&\le\delta_n^{1/2}n^{-1/4},
\end{align*}
for $\delta_n=o(1)$ and a constant $q>2$. Furthermore, we impose the following regularity conditions for all $n\ge 3$, $P\in\mathcal{P}$ and $q>2$: Given a random subset $I$ of $[n]$ of size $n/K$ with $I^c$ denoting its complement, the conditional mean estimator $\hat{\eta}=\hat{\eta}((W_i)_{i\in I^c})$ satisfies $\|\hat{\eta}-\eta_{0}\|_{P,2q}\le C$,  $\|\hat{\eta}-\eta_{0}\|_{P,4}\le \delta_n$, and $\|\hat{\eta}-\eta_{0}\|_{P,2}\le\delta_n^{1/2}n^{-1/4}$ with $P$-probability not less than $1-o(1)$.

Assumption 3.2(a) of \cite{doubleML} holds by construction of the set $\mathcal{T}_n$ and our regularity conditions. We proceed by verifying Assumption 3.2(b). For $q>2$, we have
\begin{align*}
\sup_{\eta\in\mathcal{T}_n}E[|\phi_2(V,\theta_0,\eta)|^q]^{1/q}&=\sup_{\eta\in\mathcal{T}_n}E\left[\left((\eta_1(V)-\eta_2(V))^2-\theta_0+\zeta\right)^q\right]^{1/q}\\
&\le \sup_{\eta\in\mathcal{T}_n}\|(\eta_1(V)-\eta_2(V))^2\|_{P,q}+|\theta_0|+\|\zeta\|_{P,q}\\
&\lesssim \sup_{\eta\in\mathcal{T}_n}\|(\eta_1(V)-\eta_2(V))^2\|_{P,q}+C
\end{align*}
due to $\|\zeta\|_{P,q}<C$ with
\begin{align*}
&\quad\sup_{\eta\in\mathcal{T}_n}\|(\eta_1(V)-\eta_2(V))^2\|_{P,q}\\
&=\sup_{\eta\in\mathcal{T}_n}\|\eta_1(V)-\eta_2(V)\|_{P,2q}^{2}\\
&\le\sup_{\eta\in\mathcal{T}_n}\|(\eta_1-\eta_{1,0})(V)+(\eta_{1,0})(V)-\eta_{2,0})(V))+(\eta_{2,0}-\eta_2)(V)\|_{P,2q}^2\\
&\le \sup_{\eta\in\mathcal{T}_n}\left(\|(\eta_1-\eta_{1,0})(V)\|_{P,2q}+\|(\eta_{2,0}-\eta_2)(V)\|_{P,2q}\right)^2\\
&\le C
\end{align*}
by construction of set $\mathcal{T}_n$. Furthermore, we have
\begin{align*}
\sup_{\eta\in\mathcal{T}_n}E[|\phi_2^a(V,\theta_0,\eta)|^q]^{1/q}&=1.
\end{align*}
Next, we verify Assumption 3.2(c). It holds that
\begin{align*}
\sup_{\eta\in\mathcal{T}_n}|E[\phi_2^a(V,\theta_0,\eta)-\phi_2^a(V,\theta_0,\eta_0)]|=|(-1)-(-1)|=0.
\end{align*}
In addition, we have
\begin{align*}
&\quad\sup_{\eta\in\mathcal{T}_n}E[|\phi_2(V,\theta_0,\eta)-\phi_2(V,\theta_0,\eta_0)|^2]^{1/2}\\
&=\sup_{\eta\in\mathcal{T}_n}E\left[\left((\eta_1(V)-\eta_2(V))^2-(\eta_{1,0}(V)-\eta_{2,0}(V))^2\right)^2\right]^{1/2}\\
&=\sup_{\eta\in\mathcal{T}_n}E\bigg[\left((\eta_{1,0}(V)-\eta_{2,0}(V))+(\eta_1(V)-\eta_2(V))\right)^2\\
&\quad\left((\eta_{1,0}(V)-\eta_{2,0}(V))-(\eta_1(V)-\eta_2(V))\right)^2\bigg]^{1/2}\\
&\le\sup_{\eta\in\mathcal{T}_n}E\left[((\eta_{1,0}(V)-\eta_{2,0}(V))+(\eta_1(V)-\eta_2(V)))^4\right]^{1/4}\\
&\quad E\left[((\eta_{1,0}(V)-\eta_{2,0}(V))-(\eta_1(V)-\eta_2(V)))^4\right]^{1/4}\\
&\le C\sup_{\eta\in\mathcal{T}_n}(E\left[((\eta_{1,0}(V)-\eta_1(V))+(\eta_2(V)-\eta_{2,0}(V)))^4\right]^{1/4}\\
&\le 2C\sup_{\eta\in\mathcal{T}_n}\|\eta_{0}-\eta\|_{P,4}\lesssim \delta_n.
\end{align*}
Moreover, for all $r\in (0,1)$, it holds
\begin{align*}
&\quad\sup_{\eta\in\mathcal{T}_n}\partial_r^2E[\phi_2(V,\theta_0,\eta_0 + r(\eta-\eta_0)]\\
&=\sup_{\eta\in\mathcal{T}_n}E\left[\partial_r^2\phi_2(V,\theta_0,\eta_0 + r(\eta-\eta_0))\right]\\
&=\sup_{\eta\in\mathcal{T}_n}E\left[\partial_r^2(\eta_{1,0}(V)+r(\eta_1(V)-\eta_{1,0}(V))-(\eta_{2,0}(V)+r(\eta_2(V)-\eta_{2,0}(V))))^2\right]\\
&=\sup_{\eta\in\mathcal{T}_n}E\bigg[\partial_r 2(\eta_{1,0}(V)+r(\eta_1(V)-\eta_{1,0}(V))-(\eta_{2,0}(V)+r(\eta_2(V)-\eta_{2,0}(V))))\\
&\quad((\eta_1(V)-\eta_{1,0}(V))-(\eta_2(V)-\eta_{2,0}(V)))\bigg]\\
&=\sup_{\eta\in\mathcal{T}_n}E\left[2((\eta_1(V)-\eta_{1,0}(V))-(\eta_2(V)-\eta_{2,0}(V)))^2\right]\\
&\le C \sup_{\eta\in\mathcal{T}_n}\|\eta-\eta_0\|_{P,2}^2\\
&\le C(\delta_n^{1/2}n^{-1/4})^2=C\delta_nn^{-1/2}.
\end{align*}
Finally, we show that the variance of the score $\phi_2$ is non-degenerate:
\begin{align*}
E[\phi_2(V,\theta_0,\eta_0)^2]&=E[( (\eta_{1,0}(V)-\eta_{2,0}(V))^2-\theta_0+\zeta)^2]\\
&=E[((\eta_{1,0}(V)-\eta_{2,0}(V))^2-\theta_0)^2]+E[\zeta^2]\\
&\ge E[\zeta^2]=\sigma_{\zeta}^2
\end{align*}
since the variance $\sigma_{\zeta}^2$ of $\zeta$ is chosen to be bounded from below. This completes the proof of Neyman orthogonality and asymptotic normality of our test under the null hypothesis.
}
\end{document}